\RequirePackage{snapshot}
\documentclass[%
reprint,
amsmath,amssymb,
aps,
prx,
nobalancelastpage,
onecolumn,
nofootinbib
]{revtex4-2}

\usepackage{graphicx}
\usepackage{dcolumn}
\usepackage{bm}
\usepackage[caption=false]{subfig}
\usepackage{url}
\usepackage{overpic}
\usepackage{xcolor}
\usepackage{enumitem}
\usepackage{placeins}
\usepackage{booktabs}
\usepackage{makecell}

\DeclareMathAlphabet{\mymathbb}{U}{bbold}{m}{n} 

\makeatletter
\def\bibsection{%
   \par
   \begingroup
    \baselineskip26\p@
    \bib@device{\hsize}{72\p@}%
   \endgroup
   \nobreak\@nobreaktrue
   \addvspace{19\p@}%
  }%
\makeatother

\begin{document}

\newcommand{\ket}[1]{| #1 \rangle}
\newcommand{\bra}[1]{\langle #1 |}
\newcommand{\1}{\mymathbb{1}} 

\title{Combatting noise in near-term quantum data centres} 

\author{K. Campbell}
\email{kennycampbell1@btinternet.com}
\author{A. Lawey}
\author{M. Razavi}
\email{m.razavi@leeds.ac.uk}
\affiliation{School of Electronic and Electrical Engineering, University of Leeds, Leeds, LS29JT, United Kingdom}

\begin{abstract}
We analyse the performance of different error handling methods in the quantum data centre paradigm of distributed quantum computing. We compare the impact of quantum error detection, using the three-qubit repetition code and the [[4, 1, 2]] Leung-Nielsen-Chuang-Yamamoto code, on remote gates with that of conventional entanglement distillation techniques. Detailed classical simulation is used to obtain results for realistic near-term hardware.
% We propose the use of localised quantum error detection encoding as a means of mitigating entanglement errors in the quantum data centre paradigm of distributed quantum computing. Specific implementations of such localised encodings are introduced for the three-qubit repetition code and the [[4, 1, 2]] Leung-Nielsen-Chuang-Yamamoto code \cite{412_subcode}. The performance of these schemes is evaluated using classical simulation and compared to BBPSSW and DEJMPS entanglement distillation.  
\end{abstract}

\maketitle

\section{Introduction} \label{sec:introduction}

Distributed quantum computing is a promising strategy for scaling up quantum computers. By using a distributed architecture, it is possible to retain the high quality of local gates and physical qubits needed for fault-tolerant computing while circumventing the challenges associated with scaling up monolithic, single-processor, devices. However, new challenges arise in dealing with noisy entanglement between quantum processing units (QPUs) \cite{our_first_paper}. Previous work \cite{our_first_paper} indicates that entanglement errors dominate other forms of error, even when QPUs are in close proximity to each other, making entanglement errors a severe bottleneck to the feasibility of distributed quantum computing. In this work, we propose and evaluate various near-term strategies for mitigating the impact of entanglement noise over distributed quantum computers. Our focus will be on so-called quantum data centres (QDCs) \cite{our_first_paper}, which have entangling connections between QPUs in the vicinity of each other. 

A QDC consists of multiple spatially separated QPUs that each contain qubits and the control systems needed to apply unitary gate operations and measurements on them. QPUs are close enough together to share the same warehouse, with typical separations being of the order of metres or less. These QPUs are networked together to potentially enable larger scale computation. We note that alternative, less prescriptive, definitions of QDCs also exist \cite{altQDCdef,   liu_quantum_2024}, but these are not the focus of this work.

% An example set-up for an entire QDC is shown in Fig. \ref{fig:QDC}(a), while Fig. \ref{fig:QDC}(b) shows a zoomed in depiction of neighbouring QPUs and the link between them. %
% \begin{figure}
%     \centering
%     \begin{overpic}[scale=1.0]{fig1a_large_QDC_diagram.pdf}%
%         \put(-15, 80){(a)}    
%     \end{overpic}\hspace{4em}%
%     \begin{overpic}[scale=1.0]{fig1b_QDC_diagram.pdf}
%         \put(18, 40){(b)}
%     \end{overpic}
%     \caption{(a) A quantum data centre. (b) The linking of two QPUs in a QDC. Communication qubits are shown in orange with a dotted border. The communication qubits on each QPU are linked together via a quantum-classical connection through which entanglement is distributed. Also present on each QPU are processing qubits, shown in blue with a solid border. Processing qubits are  manipulated with local gates, and are used in much the same way as qubits on a monolithic device. The number of communication qubits is typically more limited than the number of processing qubits. This figure is taken from our previous work \cite{our_first_paper}.}
%     \label{fig:QDC}
% \end{figure}%

Many of the challenges of implementing QDCs coalesce around one central issue: what happens when we wish to enact a two-qubit gate on qubits hosted by different QPUs? We call such a gate a remote gate. Remote gates involve several local, intra-QPU, operations, which will be imperfect and add noise to the system; classical communication, which adds latency; and, most damagingly of all \cite{our_first_paper}, inter-QPU entanglement, or ebit, errors. Unmanaged, these errors severely inhibit the feasibility of QDCs \cite{our_first_paper} and so handling them is essential.

We consider two broad categories of error handling: quantum error detection (QED) and entanglement distillation, in the context of a remote CNOT gate. QED is closely related to the more powerful concept of quantum error correction \cite{QuantumErrorCorrectionIntroductoryGuide}, which can, in principle, eradicate errors from a quantum computer. However, the stringent requirements needed to implement fault-tolerant, error-corrected qubits of sufficient quality and quantity for useful applications exceed the capabilities of current hardware. It is far easier to detect errors and post-select results in which no errors are detected. It is also possible to do entanglement distillation, which uses local operations and classical communication to iteratively convert multiple noisy ebits into fewer, higher quality ebits \cite{BBPSSW, DEJMPS}. As with QED, entanglement distillation must be repeated until success and so has an associated time cost, leading to additional decoherence. 

In our implementation of the remote CNOT gate using QED, instead of applying QED to ebits, we apply QED to the control qubit, in an arbitrary state, which we will then teleport. We encode the qubit's state in a stabiliser QED code \cite{NielsenChuang} prior to teleportation and detect errors directly on the teleported state. In theory, our method has the advantage of also detecting errors caused by the local gates and measurements involved in the teleportation process but has the disadvantage that any detected errors require the entire algorithm in which the teleportation occurs to be restarted.

In contrast to much of the previous work on QED in the computing context \cite{gottesman2016FTwithED, Roffe_2018, Linke_2017, Harper_2019, ED_IBM5Q, Knill_2005}, we do not assume (QED) encoding is retained throughout the entire algorithm, which would require many logical gates and qubits. Instead, we propose the localised application of QED to  remote gates, in which QED encoding is performed immediately prior to a remote gate and decoding is performed during the remote gate. The most similar work to ours is Ref. \cite{PauliChecksQuantumNetworks} but they consider detecting errors only on ebits and not the arbitrary unknown state of processing qubits.

In our implementation of the remote CNOT gate using entanglement distillation, we extend previous work \cite{Tele4DQC, EntDist4DQC, siddhu2025basicdistillationrealisticnoise} by considering arbitrary input states, rather than assuming a known state is teleported. We also explicitly consider an entire remote CNOT gate rather than only the teleportation part \cite{Tele4DQC, EntDist4DQC} or only the distribution of ebits \cite{siddhu2025basicdistillationrealisticnoise}. This impacts the results if local errors are considered. 

We make several additions to the existing literature. Our main contributions are as follows:
\begin{itemize}
    \item We propose the notion of localised QED encoding as a means to improve the fidelity of remote gates in the QDC setting. 
    \item We perform a circuit-level analysis of various QED schemes.
    \item We compare the relative performance of QED and entanglement distillation for near-term QDCs.
\end{itemize}

The remainder of the paper is structured as follows. In Sec. \ref{sec:system_description}, we introduce the schemes analysed in this paper.  The error analysis conducted on these schemes is discussed in Sec. \ref{sec:error_analysis}. The results are then presented in Sec. \ref{sec:results}. Finally, we conclude our discussion in Sec. \ref{sec:conclusion}.

\section{System description}\label{sec:system_description}

As CNOT and single qubit gates are sufficient for universal quantum computing \cite{NielsenChuang}, we restrict ourselves to the study of remote CNOT gates for concreteness. We first consider the basic, unencoded remote gate in Sec. \ref{sec:unencoded_case} and then detail the additions that we make to it using QED, in Sec. \ref{sec:qed_schemes}, and entanglement distillation, in Sec. \ref{sec:ent_dist_schemes}.

\subsection{Unencoded case}\label{sec:unencoded_case}

\begin{figure}
    \centering
    \includegraphics[width=0.5\linewidth]{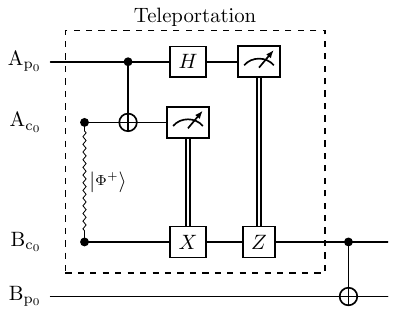}
    \caption{A remote CNOT gate between qubits $\mathrm{A_{p_0}}$ and $\mathrm{B_{p_0}}$ implemented using the 1TP protocol. Zigzag lines indicate the distribution of an ebit, which, ideally, in the absence of noise, has the state $\ket{\Phi^+} = \frac{1}{\sqrt{2}} ( \ket{00} + \ket{11})$. Double lines indicate classical communication. The subscript $\mathrm{c}_i$ indicates a communication qubit with index $i$ and $\mathrm{p}_i$ is a processing qubit with index $i$.}
    \label{fig:1tp_circuit_diagram}
\end{figure}

We use a simple scheme, which we call single teleportation, 1TP \cite{our_first_paper}, to implement remote gates, as shown in Fig. \ref{fig:1tp_circuit_diagram}. %
To implement a 1TP remote CNOT, the control qubit, $\mathrm{A_{p_0}}$, is teleported \cite{teleportationProposal, firstExperimentalTeleportation} to ${\mathrm{B_{c_0}}}$ on the QPU containing the target qubit, $\mathrm{B_{p_0}}$, and then the CNOT gate is conducted locally using $\mathrm{B_{c_0}}$ as the control qubit and $\mathrm{B_{p_0}}$ as the target qubit. Three different types of qubit are used: communication qubits, $\mathrm{A_{c_0}}$ and $\mathrm{B_{c_0}}$, processing qubits, $\mathrm{A_{p_0}}$ and $\mathrm{B_{p_0}}$, and flying qubits, which are not shown explicitly in Fig. \ref{fig:1tp_circuit_diagram}. The flying qubits, which are typically photons, are used to distribute entanglement between the communication qubits on each QPU. The processing qubits, which may or may not be physically distinct from the communication qubits, are used much like qubits in a monolithic setting and do not interact directly with flying qubits. 

\subsection{QED schemes}\label{sec:qed_schemes}

\begin{figure}
    \subfloat[FC-1TP]{\begin{overpic}[width=0.53\textwidth]{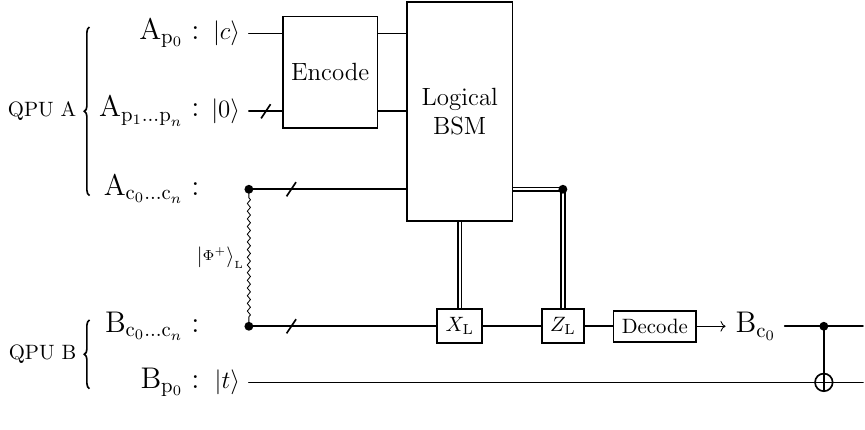}
        \put(60, 38){$\rightarrow$}
        \put(65,31){\frame{\includegraphics[width=0.2\textwidth]{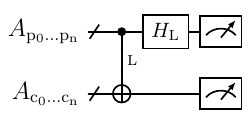}}}
    \end{overpic}}\hfill
    \subfloat[PC-1TP]{\includegraphics[width=0.44\textwidth]{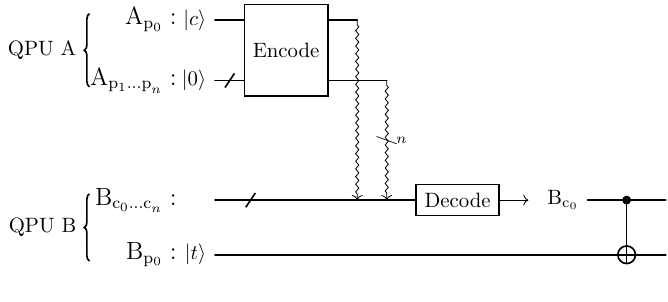}}
    \caption{1TP remote CNOT gate using (a) fully-coded 1TP (FC-1TP) and (b) partially-coded 1TP (PC-1TP). The subscript L refers to a logical version of the operation or state. We omit the L for the measurements in (a) but both measurements are logical measurements. Double lines refer to classical communication, squiggly lines to the distribution of an ebit and squiggly arrows to quantum teleportations from the tail of the arrow to its head. $\ket{c}$ and $\ket{t}$ are arbitrary quantum states, which we assume to be separable from each other and pure in this work. If an error is detected during the decoding step then the final CNOT is not done and the entire process must be restarted.}
    \label{fig:qed_schemes}
\end{figure}

We consider two broad ways of integrating QED into a 1TP remote CNOT gate, as shown in Fig. \ref{fig:qed_schemes}. %
The first scheme, shown in Fig. \ref{fig:qed_schemes}(a), is referred to as fully-coded 1TP (FC-1TP) and ideally works as follows:
\begin{enumerate}
    \item We first distribute a logical ebit, of the form $\ket{\Phi^+} = \frac{1}{\sqrt{2}}(\ket{00}_{\mathrm{L}} + \ket{11}_{\mathrm{L}})$, between both QPUs involved in the remote gate, where $\ket{0}_{\mathrm{L}}$ and  $\ket{1}_{\mathrm{L}}$ are the logical computational basis eigenstates encoded in the QED code of choice. This is done, using the method discussed in Ref. \cite{QuantumRepeaterWithEncoding} for encoded quantum repeaters, as follows:
    \begin{enumerate}
        \item Qubits $A_{c_0...c_n}$ are initially encoded in the logical state $\ket{+}_{\mathrm{L}} = \frac{1}{\sqrt{2}}(\ket{0}_{\mathrm{L}} + \ket{1}_{\mathrm{L}})$ and qubits $B_{c_0...c_n}$ are encoded in the logical state $\ket{0}_{\mathrm{L}}$.
        \item A logical remote CNOT gate is done between qubits $A_{c_0...c_n}$ and $B_{c_0...c_n}$. In our case, we only consider the repetition code for the first scheme and so $\ket{0}_{\mathrm{L}} = \ket{000}$ and $\ket{1}_{\mathrm{L}} = \ket{111}$ and the logical remote CNOT gate is done transversally using additional communication qubits, not shown in Fig. \ref{fig:qed_schemes}(a), to implement each of the physical remote gates. However, it is easy to generalise the same concepts to different QED codes.
    \end{enumerate}
    \item Then, using $A_{p_1...p_n}$, we apply local gates to encode qubit $\mathrm{A_{p_0}}$.
    \item After that, a logical BSM is applied, as shown in the box in the top-right corner of Fig. \ref{fig:qed_schemes}(a). Firstly, a logical CNOT is conducted between $A_{p_0...p_n}$ and $A_{c_0...c_n}$. Then, a logical X-basis measurement is done on $A_{p_0...p_n}$ and a logical Z-basis measurement is done on $A_{c_0...c_n}$. 
    \item The results of the measurements are sent from QPU A to QPU B and the local X and Z operations needed to complete the teleportation protocol are done on $B_{c_0...c_n}$.
    \item Finally, decoding is done and a physical CNOT gate is done between $B_{\mathrm{c_0}}$ and $B_{\mathrm{p_0}}$. If errors are detected during decoding or the logical BSM, the CNOT gate is not done and the entire process is repeated from the beginning. 
\end{enumerate} 

In the second scheme, shown in Fig. \ref{fig:qed_schemes}(b), logical ebits are not used. Instead, $\mathrm{A_{p_0}}$ is encoded into a QED code, using $A_{p_1...p_n}$, as before, and then qubits $A_{p_0...p_n}$ are individually teleported to $B_{c_0...c_n}$, using additional communication qubits and ebits not shown in Fig. \ref{fig:qed_schemes}. In our case, we assume that the teleportations are done sequentially and so only a single communication qubit is required on QPU A. After the teleportations, decoding is done, as in FC-1TP, and a local, unencoded, CNOT gate is done locally between $B_{\mathrm{c_0}}$ and $B_{\mathrm{p_0}}$. We refer to this scheme as partially-coded 1TP (PC-1TP).

PC-1TP has the advantage of avoiding the need for any logical operations. This allows it to be used with a larger class of codes, and makes the implementations of arbitrary codes simpler.  This makes it trivial to extend PC-1TP to any QED code without the need to worry about the fact that different QED codes use different logical gate implementations. 

\begin{figure}
    \centering
    \subfloat[3QRC]{\includegraphics[scale=0.6]{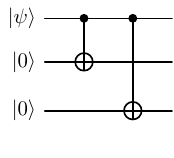}}\hspace{3em}%
    \subfloat[4QED]{\includegraphics[scale=0.6]{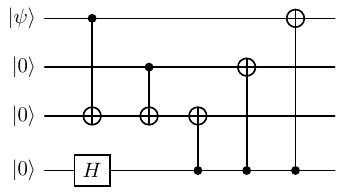}}\hspace{3em}%
    \subfloat[SS]{\includegraphics[scale=0.6]{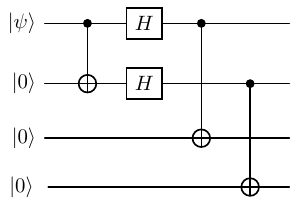}}
    % \begin{overpic}[scale=0.6]{fig3a_repetition_code_encoder_diagram.pdf}
    %     \put(-20, 55){(a)}
    % \end{overpic}\hspace{3em}%
    % \begin{overpic}[scale=0.6]{fig3b_four_qubit_ED_diagram.pdf}
    %     \put(-5, 55){(b)}
    % \end{overpic}\hspace{3em}%
    % \begin{overpic}[scale=0.6]{fig3c_short_shor_diagram.pdf}
    %     \put(-10, 55){(c)}        
    % \end{overpic}
    \caption{Encoding circuits for (a) the three-qubit repetition code, (b) the LNCY code \cite{412_subcode} code using the encoder from \cite{QuantumErrorCorrectionIntroductoryGuide}, (c) a simpler encoding of the LNCY. $\ket{\psi}$ is an arbitrary quantum state.}
    \label{fig:encoding_circuits}
\end{figure}

Figure \ref{fig:encoding_circuits} %
shows the encoding circuits for the specific codes used in this paper.
Figure \ref{fig:encoding_circuits}(a) encodes an arbitrary quantum state into the three-qubit repetition code, which we shall refer to as 3QRC. This has the logical encoding $\ket{0}_{\mathrm{L}} = \ket{000}$ and $\ket{1}_{\mathrm{L}} = \ket{111}$ on the basis states. An arbitrary state, $\ket{\psi} = \alpha \ket{0} + \beta \ket{1}$, where $\alpha, \, \beta \in [0, 1]$ are normalised such that $|\alpha|^2 + |\beta|^2 = 1$, is encoded as $\ket{\psi}_{\mathrm{L}} = \alpha \ket{000} + \beta \ket{111}$. We consider both FC-1TP and PC-1TP with 3QRC and refer to the former as FCRC and the latter as PCRC.  Figures \ref{fig:encoding_circuits}(b) and \ref{fig:encoding_circuits}(c) encode the [[4, 1, 2]] Leung-Nielsen-Chuang-Yamamoto (LNCY) \cite{412_subcode}, which, up to qubit relabelling, has  the logical eigenstates $\ket{0}_{\mathrm{L}} = \frac{1}{\sqrt{2}} (\ket{0000} + \ket{1111})$ and $\ket{1}_{\mathrm{L}} = \frac{1}{\sqrt{2}} (\ket{0101} + \ket{1010}))$.  These are the $\{\ket{00}_\mathrm{L}, \, \ket{10}_\mathrm{L}\}$ logical basis states of the four-qubit code \cite{Linke_2017}. The encoding circuit shown in Fig. \ref{fig:encoding_circuits}(c) is slightly more resource efficient, requiring fewer gates, but we include the encoding shown in Fig. \ref{fig:encoding_circuits}(b) too, for the sake of comparison. We refer to the schemes shown in Figs. \ref{fig:encoding_circuits}(b) and \ref{fig:encoding_circuits}(c) as 4QED and SS, for shortened Shor\footnote{Both of the [[4, 1, 2]] codes used are equivalent if one of the qubits in the first scheme is used as a gauge qubit, meaning it is essentially ignored, and both codes can be thought of as a shortened version of the Shor code \cite{9qubitshorErrorCorrection}, but, in the interests of telling them apart, we attach this name only to one of the codes.}, respectively. For simplicity, we consider only PC-1TP for both 4QED and SS.

\begin{figure}
    \centering
    \subfloat[3QRC]{\includegraphics[scale=0.6]{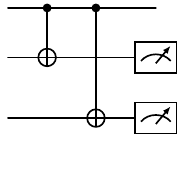}}\hspace{3em}%
    \subfloat[4QED]{\includegraphics[scale=0.6]{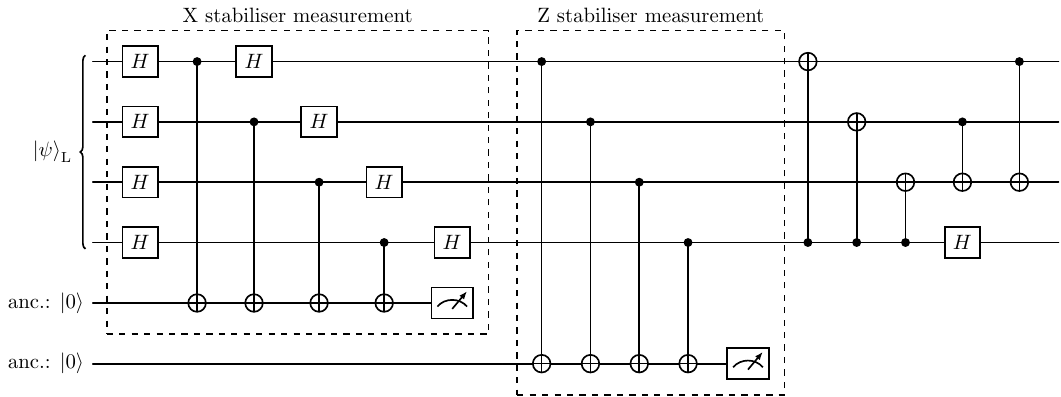}}\hspace{3em}%
    \subfloat[SS]{\includegraphics[scale=0.6]{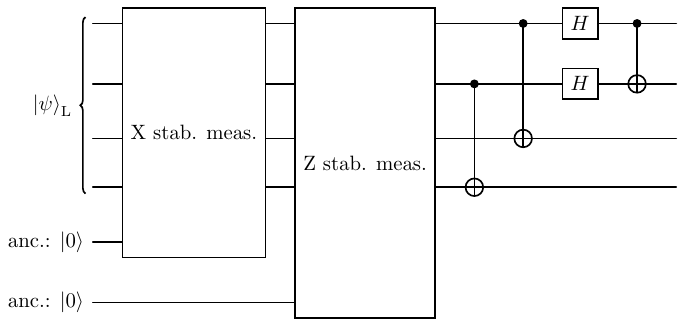}}
    % \begin{overpic}[scale=0.6]{fig4a_repetition_code_decoder_diagram.pdf}
    %     \put(-10, 55){(a)}
    % \end{overpic}\hspace{3em}%
    % \begin{overpic}[scale=0.6]{fig4b_four_qubit_ED_decoder_diagram.pdf}
    %     \put(0, 35){(b)}        
    % \end{overpic}\\ \vspace{1em}%
    % \begin{overpic}[scale=0.6]{fig4c_short_shor_decoder_diagram.pdf}
    %     \put(0, 40){(c)}
    % \end{overpic}
    \caption{Decoding/error detection circuits for (a) 3QRC, (b) 4QED, and (c) SS. The X and Z stabiliser measurements indicated in (c) are identical to those explicitly shown in (b). Qubits labelled `anc.' are ancilla qubits. Everything after the Z stabiliser measurement in (b) and (c) is only done if no error was detected during the X and Z stabiliser measurements. If an error is detected, the result is discarded and the entire operation must be discarded.}
    \label{fig:decoding_circuits}
\end{figure}

Unitary decoding of an arbitrary encoded state is done by running the encoding circuits in reverse\footnote{If the encoding gates all commute, as they do for 3QRC, then the encoding circuit can be used unchanged.}. For the 3QRC decoder, we integrate error detection into the decoder similarly to the proposal in Ref. \cite{JingAlsina2020}. For 4QED and SS, we instead detect errors prior to decoding using a stabiliser measurement \cite{QuantumErrorCorrectionIntroductoryGuide} and then run the unitary decoding circuits. This requires the use of additional, ancilla qubits. The decoding/error detecting circuits used are shown in Fig. \ref{fig:decoding_circuits}. %
The X stabiliser measurements in Figs. \ref{fig:decoding_circuits}(b) and \ref{fig:decoding_circuits}(c) detect Z errors and the Z stabiliser measurements detect X errors \cite{NielsenChuang, QuantumErrorCorrectionIntroductoryGuide}.

\subsection{Entanglement distillation schemes}\label{sec:ent_dist_schemes}

\begin{figure}
    \centering
    \subfloat[BBPSSW]{\includegraphics{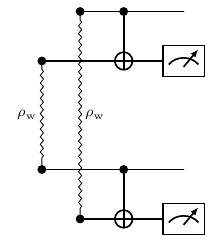}}\hspace{3em}%
    \subfloat[DEJMPS2]{\includegraphics{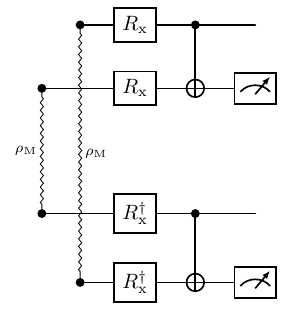}}\hspace{3em}%
    \subfloat[DEJMPS4]{\includegraphics{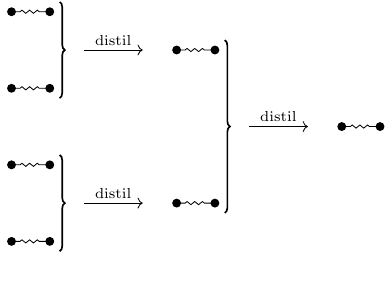}}
    % \begin{overpic}{fig5a_BBPSSW_diagram.pdf}
    %     \put(-10, 55){(a)}
    % \end{overpic}\hspace{3em}%
    % \begin{overpic}{fig5b_DEJMPS_diagram.pdf}
    %     \put(-10, 55){(b)}        
    % \end{overpic}
    \caption{The (a) BBPSSW, (b) DEJMPS and (c) general two-round entanglement distillation schemes. BBPSSW assumes a Werner state input, which can be enforced with local operations, although this is not done here, while DEJMPS allows a more general input. For both schemes, the circuit is run repeatedly until both of the measurements shown give the same result. The squiggly lines indicate the distribution of the state $\ket{\Phi^+} = \frac{1}{\sqrt{2}} ( \ket{00} + \ket{11})$, in the ideal case where no noise is present. In reality, noise will be present and for BBPSSW, it is assumed that this noise has the Werner form, as discussed in the main text. For DEJMPS, no such assumptions are made and the state $\rho_{\mathrm{M}}$ can instead be distributed, which is an arbitrary mixture of the different Bell states $\ket{\Phi^{\pm}} = \frac{1}{\sqrt{2}}(\ket{00} \pm \ket{11})$, $\ket{\Psi^{\pm} = \frac{1}{\sqrt{2}}(\ket{01} \pm \ket{10})}$. $R_x$ is a $\frac{\pi}{2}$ rotation about the $x$ axis and $R_x^{\dagger}$ is the $-\frac{\pi}{2}$ rotation. (a) and (b) show just one round of entanglement distillation here, but the fidelity can be further improved by further distilling the outcome as illustrated in (c).}
    \label{fig:entanglement_distillation_QC}
\end{figure}

To implement a remote CNOT gate with entanglement distillation, we use the unencoded scheme shown in Fig. \ref{fig:1tp_circuit_diagram} but apply entanglement distillation to the ebits. Two entanglement distillation schemes are considered, the scheme proposed in Ref. \cite{BBPSSW}, termed BBPSSW, and that of Ref. \cite{DEJMPS}, termed DEJMPS, hereafter. The schemes are shown in Figs. \ref{fig:entanglement_distillation_QC}(a) and \ref{fig:entanglement_distillation_QC}(b), respectively. %
BBPSSW is conducted by first distributing two ebits, which are assumed to be in the Werner state, defined as follows: %
\begin{equation}
    \label{eq:werner_state}
        \rho_{\rm w} =  F_{\rm w} \ket{\Phi^+} \bra{\Phi^+} 
        + \frac{1 - F_{\rm w}}{3}\left( \ket{\Phi^-} \bra{\Phi^-}  + \ket{\Psi^+} \bra{\Psi^+}
          + \ket{\Psi^-}\bra{\Psi^-} \right),
\end{equation}
where $F_{\mathrm{w}} \in [0, \, 1]$, $\ket{\Phi^{\pm}} = \frac{1}{\sqrt{2}} (\ket{00} \pm \ket{11})$, and $\ket{\Psi^{\pm}} = \frac{1}{\sqrt{2}}(\ket{01} \pm \ket{10})$. This assumption can be enforced using random single-qubit gates \cite{BBPSSW}, but, in this work, we assume that the state of the initial ebit already has the above form, as we consistently use the Werner state to model entanglement noise throughout this work.  BBPSSW also requires $F_{\mathrm{w}} \geq 0.5$. It is important to note that, although the ebits are assumed to be distributed in the Werner state, they are also subjected to additional memory depolarisation after successful distribution has occurred. This and the other noise models considered are discussed in Sec. \ref{sec:noise_models}.

The second scheme, DEJMPS, assumes that the ebits have the state 
\begin{equation}\label{eq:rho_M}
    \rho_{\mathrm{M}} = A \ket{\Phi^+}\bra{\Phi^+} + B \ket{\Psi^-}\bra{\Psi^-} + C \ket{\Psi^+}\bra{\Psi^+} + D \ket{\Phi^-}\bra{\Phi^-},
\end{equation}
where $A, \, B,\, C,\, D \in [0, 1]$, $A + B + C + D=1$, and $A\geq 0.5$. However, for consistency with our error analysis of the QED schemes, we again assume that the initial, pre-distillation ebits have the Werner state in Eq. \eqref{eq:werner_state}.

After the entanglement distribution step, DEJMPS applies the operation $R_{\mathrm{X}}\otimes R_{\mathrm{X}}^{\dagger}$, to each ebit, where $R_{\mathrm{X}}$ is a $+\frac{\pi}{2}$ rotation about the $x$ axis. This swaps B and D in Eq. \eqref{eq:rho_M}. After that, the same local operations and measurements as BBPSSW are used. 

Both schemes start with two pairs of entangled states entangled between two QPUs, apply CNOT gates locally on their share of the entangled states, and then measure the target qubits. The measurement results are compared and if they do not have the same value then the unmeasured qubits are discarded. This process distinguishes between the $\ket{\Phi^{\pm}}$ and $\ket{\Psi^{\pm}}$ states \cite{BBPSSW}. The entire process is repeated until both measurement results are the same.

It is possible to improve the fidelity by nesting the entanglement distillation process so that already distilled ebits are distilled again. This is shown in Fig. \ref{fig:entanglement_distillation_QC}(c). DEJMPS is known to perform better than BBPSSW when multiple nested rounds of entanglement distillation are done \cite{DEJMPS} and so we consider nesting only for DEJMPS. In our simulated implementations of the nested DEJMPS protocol, discussed in Sec. \ref{sec:discrete_event_simulation}, four ebits are initially distributed sequentially, waiting for one distribution to finish before starting the next. This is not optimal and introduces asymmetries in the fidelities of the ebits, as some ebits spend longer depolarising on the memory than others. Clearly, parallelisation could be done here, however for ease of simulation, we have not done so. Moreover, exact parallelisation would be challenging due to the probabilistic nature of entanglement distillation and so we would expect different ebits to experience different amounts of memory depolarisation in reality too. Our results capture this behaviour but the fully sequential operation represents a larger latency and memory depolarisation than would be expected in reality. As such, our results represent a conservative lower bound on performance. After the ebits are distributed, they are distilled into two higher-fidelity ebits. Those two distilled ebits are then distilled again to make a single ebit. If either distillation fails, we repeat the entire process again, creating four ebits and distilling them until the entire protocol succeeds\footnote{If one of the distillations is successful and the other fails, we could retain the successfully distilled ebit and discard only the unsuccessfully distilled one. This would arguably be more efficient but risks memory decoherence damaging the previously distilled ebit while it waits for another ebit to be successfully distilled. To avoid such complications, we discard both pairs but note that this may be sub-optimal.}. We refer to this nested protocol as DEJMPS4 to distinguish it from the single-round case where we start with two ebits and distill them into a single ebit, which we call DEJMPS2.

\section{Error analysis}\label{sec:error_analysis}

We subject the various implementations of the remote CNOT gate discussed in Sec. \ref{sec:system_description} to various types of error. We do this using discrete-event simulation, which is discussed in Sec. \ref{sec:discrete_event_simulation}. The noise models used are discussed in Sec. \ref{sec:noise_models}. In Secs. \ref{sec:average_gate_fidelity} and \ref{sec:latency}, we introduce two of the key figures of merit that we consider, average gate fidelity and latency, respectively.

\subsection{Discrete-event simulation}\label{sec:discrete_event_simulation}

The majority of the results in this work are taken using discrete-event simulation with NetSquid \cite{netsquid} and the related dqc\_simulator package we developed \cite{MySimGithub}. This means that time, $t$, does not progress continuously but is instead a discrete variable, which increases in increments only when something, an event, happens. For example, a classical message arriving at the receiver would be an event and only when the message arrives would we apply the memory decoherence accrued when waiting for that message. All classical and quantum operations we consider have an associated duration, and  when the operation finishes, we increment $t$ by that amount. Some operations also have associated noise models, which are implemented using quantum channels. Further details on the channels used are provided in Sec. \ref{sec:noise_models}. The quantum state of the system is tracked by grouping together all qubits that are entangled with each other and tracking the density matrix for each group of qubits after a given event, re-assigning qubits to new groups to represent changes in entanglement, where appropriate. 

Some abstractions and constraints are used for convenience of simulation. Firstly, ebits are generated at a black box central source in the Werner state, given by Eq. \eqref{eq:werner_state} and the constituent qubits travel to the QPUs. Upon arrival, the qubits are treated as communication qubits and we do not explicitly model the transduction between communication qubits and photons, although it is accounted for in the rate at which ebits are generated. The noise generated by the various subprocesses that would occur in real experiments is amalgamated into Eq. \eqref{eq:werner_state}. Similarly, the latency of the entire process is amalgamated into a single parameter, $t_{\mathrm{dist}}$, representing the entanglement distribution time.  Secondly, the remote gates or individual teleportations used by the FC-1TP and PC-1TP, respectively are done sequentially, due to limitations on the simple scheduler used. For both entanglement distillation schemes, ebit distribution is again done sequentially. With more advanced scheduling, parallelisation would be possible for ebit distribution, remote gates and teleportations. That said, even in the parallelised regime, the probabilistic nature of entanglement distillation means that we will not obtain all ebits at once and what we do conservatively upper bounds the impact of memory depolarisation that could occur because of this. Finally, we neglect the classical processing time needed for classical logic. For example, the time taken to interpret classical messages or recognise that ebits have arrived at a QPU is neglected. 

\subsection{Noise models}\label{sec:noise_models}

All ebits considered in this work are assumed to be initially distributed in the Werner state defined in Eq. \eqref{eq:werner_state}. In addition to this, we model other sources of noise in our system using several variants of the depolarising noise channel \cite{NielsenChuang}.

All qubits, including the two qubits comprising an ebit, undergo memory depolarisation, which is acted on qubits individually as time passes. This is done using the channel %
\begin{equation} \label{eq:mem_depol_channel}
    \rho_{\mathrm{in}} \rightarrow e^{- \Delta t \: r} \rho_{\mathrm{in}} + \frac{(1-e^{- \Delta t \: r})}{3} (X_i \rho_{\mathrm{in}} X_i + Y_i \rho_{\mathrm{in}} Y_i + Z_i \rho_{\mathrm{in}} Z_i),
\end{equation}
where $i$ is the qubit being acted on; $\Delta t$ is the time since the last event; $X_i$, $Y_i$, and $Z_i$ are the Pauli operations acting on qubit $i$; and $r$ is the memory depolarisation rate, which is a parameter of the hardware. As detailed in Sec. \ref{sec:discrete_event_simulation}, we use event-based simulation, and so any physical process, such as a gate or the sending of a classical message has an associated duration. After each event, we apply Eq. \eqref{eq:mem_depol_channel} to all relevant qubits.

In addition to the memory depolarisation imposed on qubits while they idle on each QPU, we also consider noise induced by imperfect operations. First, consider single-qubit gates. For a multi-qubit system in state $\rho_{\mathrm{in}}$, acting the unitary operation $U_i$ on qubit $i$ results in the following transformation: %
\begin{equation} \label{eq:single_qubit_gate_error}
    \rho_{\rm in} \rightarrow (1 - \epsilon_{\mathrm{sg}}) U_{i} \rho_{\rm in} U_{i}^{\dagger} + 
    \frac{\epsilon_{\mathrm{sg}}}{2} \mathrm{Tr}_{i}(\rho_{\mathrm{in}}) \otimes \1_{i},
\end{equation}
where $\epsilon_{\mathrm{sg}}$ is the probability of error occurring for single-qubit gates, $\mathrm{Tr}_{i}$ is the partial trace with respect to qubit $i$, and $\1_{i}$ is its corresponding identity matrix.

For a two-qubit gate, $U_{i, \, j}$, acting on qubits $i$ and $j$, the channel has the following form \cite{imperfectRepeaters}: %
\begin{equation} \label{eq:depol_2_qubit_gate_error}
    \rho_{\rm in} \rightarrow (1 - \epsilon_{\mathrm{tg}}) U_{i, \, j} \rho_{\rm in} U_{i, \, j}^{\dagger} + 
    \frac{\epsilon_{\mathrm{tg}}}{4} \mathrm{Tr}_{i, \,j}(\rho_{\mathrm{in}}) \otimes \1_{i, \, j},
\end{equation}
where $\epsilon_{\mathrm{tg}}$ is the probability of error for two-qubit gates, $\mathrm{Tr}_{i, \, j}$ is the partial trace over qubits $i$ and $j$, and $\1_{i, \, j}$ is the identity matrix on $i$ and $j$.

Finally, we impose errors during Z-basis measurements\footnote{We construct any X-basis measurements using a Hadamard followed by a Z-basis measurement, so all of our measurements are Z-basis measurements.} on qubit $i$ using the bit-flip channel: %
\begin{equation}\label{eq:meas_error_channel}
    \rho_{\mathrm{in}} \rightarrow (1 - \epsilon_{\mathrm{m}}) \rho_{\mathrm{in}} + \epsilon_{\mathrm{m}} X_i \rho_{\mathrm{in}} X_i,
\end{equation}
where $\epsilon_{\mathrm{m}}$ is the probability of a measurement error.

\subsection{Average gate fidelity}\label{sec:average_gate_fidelity}

In order to assess the impact of the noise discussed in the previous section, we consider the average gate fidelity, $\bar{F}_{\mathrm{out}}$, as a figure of merit. Average gate fidelity is the fidelity of the output from the remote gate we consider, averaged over all possible input states. We compute the average gate fidelity in two ways: via brute force Monte Carlo sampling over the restricted space of pure separable input states and using closed form analytical results over the entire input state space, to obtain more general results. When we need to differentiate between the two methods, we call the former $\bar{F}_{\mathrm{MC}}$ and the latter $\bar{F}_{\mathrm{A}}$. $\bar{F}_{\mathrm{MC}}$ is conceptually simple and allows us to explicitly investigate the behaviour of different input states, but it is computationally expensive and sampling over arbitrary input states is prohibitively costly. Therefore, for a more general evaluation of the performance of each scheme we also consider $\bar{F}_{\mathrm{A}}$. We expect both methods to yield broadly similar results and so considering these two different methods provides some mutual validation of the results.

$\bar{F}_{\mathrm{MC}}$ is derived using Monte Carlo sampling over two independent Bloch spheres, one for the control and one for the target qubit. The derivation of $\bar{F}_{\mathrm{A}}$ is more involved and we discuss it in the remainder of this section.

Average gate fidelity is a special case of the average channel fidelity, which is defined as \cite{nielsen_simple_2002}
\begin{equation}\label{eq:average_channel_fidelity}
    \bar{F}_{\mathrm{A}} = \int d\psi \bra{\psi} \mathcal{E}(\ket{\psi}\bra{\psi})\ket{\psi},
\end{equation}
where $d\psi$ is the uniform Haar measure, normalised to give $\int d\psi = 1$, and $\mathcal{E}$ is a linear completely positive trace preserving (CPTP) map \cite{NielsenChuang}, which we shall henceforth refer to as a quantum channel. To compute the average gate fidelity we need only note that for an ideal unitary gate $U$ approximated by the noisy quantum channel $\mathcal{E}$, $U^{\dagger} \circ \mathcal{E}$ is also a quantum channel, where $\circ$ indicates composition. Therefore, average gate fidelity is simply the average channel fidelity of $U^{\dagger} \circ \mathcal{E}$, which can be expressed as \cite{nielsen_simple_2002} 
\begin{equation}
    \bar{F}_{\mathrm{A}} = \int d\psi \bra{\psi} U^{\dagger} \mathcal{E}(\ket{\psi}\bra{\psi})U\ket{\psi}.
\end{equation}

As proven in Ref. \cite{horodecki_general_1999} and more simply re-derived in Ref. \cite{nielsen_simple_2002}, Eq. \eqref{eq:average_channel_fidelity} can be conveniently re-written in terms of the entanglement fidelity, $F_{\mathrm{e}} (\mathcal{E})$. Entanglement fidelity is defined by considering a maximally entangled state, $\ket{\Omega}$, between the system of interest, $R$, with Hilbert space dimension $d$, and some $d$-dimensional system, $Q$. As discussed in Sec. 9.3 of Ref. \cite{NielsenChuang}, any correctly-sized, bipartite, maximally entangled state can be used as $\ket{\Omega}$. The entanglement fidelity is then defined as 
\begin{equation}\label{eq:entanglement_fidelity}
    F_{\mathrm{e}} = \bra{\Omega} (\1 \otimes \mathcal{E})(\ket{\Omega}\bra{\Omega})\ket{\Omega},
\end{equation}
where $\1$ is the $d \times d$ identity matrix and $\otimes$ is the tensor product. $(\1 \otimes \mathcal{E})(\ket{\Omega}\bra{\Omega})$ is commonly referred to as the Choi state of $\mathcal{E}$. The relationship between entanglement fidelity and average channel fidelity is \cite{horodecki_general_1999, nielsen_simple_2002}
\begin{equation}\label{eq:average_gate_entanglement_fidelity_relation}
    \bar{F}_{\mathrm{A}} = \frac{d F_{\mathrm{e}}(\mathcal{E}) + 1}{d+1}.
\end{equation}

\begin{figure}
    \centering
    \includegraphics[scale=1]{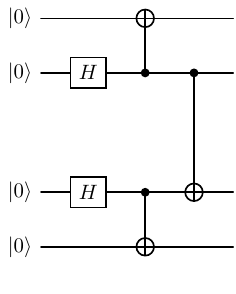}
    \caption{The circuit used to simulate the Choi state. The final CNOT gate is implemented using one of the schemes discussed in Sec. \ref{sec:system_description}. All other gates are ideal local gates. Increased spacing between quantum wires indicate that the wires belong to different QPUs.}
    \label{fig:average_gate_fidelity_setup}
\end{figure}

Equation \eqref{eq:average_gate_entanglement_fidelity_relation} is extremely powerful, as it allows average channel fidelity, and therefore average gate fidelity, to be computed without explicitly sampling over all possible input states. Instead, we can explicitly construct the Choi state in our simulation for a specific $\ket{\Omega}$ and then compute the entanglement fidelity and use this to calculate the average gate fidelity. To do this, we include ancilla qubits in our simulation and use ideal gates to maximally entangle them with the control and target qubits of the remote CNOT gate that we consider. This is depicted in Fig. \ref{fig:average_gate_fidelity_setup}. %
We note that implementing gates instantaneously can lead to scheduling issues in our discrete-event simulation and so our ``ideal" gates are actually implemented in a finite but negligible time, which causes a finite but negligible memory depolarisation of the system. The negligible time used is five orders of magnitude lower than the single-qubit gate time we consider for the noisy gates.

By calculating the fidelity between the ideal output state $\mathrm{CNOT }_{c \rightarrow t} \ket{\Omega}$ and the actual output state $(I \otimes \Lambda) \ket{\Omega}\bra{\Omega}$, where $\Lambda$ is the noisy remote gate we consider, we find the entanglement fidelity of the channel $\mathrm{CNOT} \circ \Lambda = \mathrm{CNOT}^{\dagger} \circ \Lambda$, as required to compute the average gate fidelity.

One subtlety of the arguments above is that for them to hold, our noisy remote gate, $\Lambda$, must be a quantum channel. It is not immediately obvious that this requirement is met when QED is used, due to the inclusion of post-selection in its implementation, which is not trace preserving. However, as discussed in  Appendix \ref{app:qed_channel}, we numerically find the success probability of post-selection to be approximately constant with respect to input state and so trace preservation is recovered if, as is the case here, we are only interested in the fidelity of successfully post-selected states. For the unencoded case, we use only unitary gates and quantum teleportation, which are well established quantum channels and so their composition is a quantum channel. Similarly, for entanglement distillation, post-selection is done only on the states of ancilla qubits and so linearity of the map considered is unaffected. It is also worth noting that we average the $\bar{F}_{\mathrm{A}}$ results over multiple shots to account for fluctuations in latency, which are discussed in Sec. \ref{sec:latency}, and affect the memory depolarisation. 

\subsection{Latency}\label{sec:latency}

As well as the average gate fidelity, discussed in the previous section, it is also important to consider the associated latency cost of QED and entanglement distillation. Average gate fidelity indicates the quality of a given gate but the latency of remote gates will impact the memory depolarisation of the rest of the qubits in a larger circuit and is also an important consideration to ensure that the theoretical advantage of distributed quantum computing can be converted to a practical advantage.

For a given simulated experimental shot, the latency can be extracted directly from the discrete-event simulation, which tracks the time at all points in the simulation. In our case, we take the start time to be the time after our input maximally entangled state, $\ket{\Omega}$, is generated and the end time to be when all operations are finished. The latency of a shot is simply the difference between the end time and the start time. 

The conversion of the shot latency into an overall latency then proceeds differently for the unencoded remote gate and the entanglement distillation schemes compared to the QED schemes. For the unencoded remote gate and the remote gate with entanglement distillation, we simply average over multiple experimental shots. Although, our discrete-event simulations use the density matrix formalism, which deterministically encodes the entire state, with all possibilities accounted for, the measurements in the circuit do probabilistically collapse to binary values. As such the  measurement-dependent Pauli corrections used by the simulation are applied probabilistically, changing the latency. We physically apply the Pauli corrections rather than using Pauli frame tracking because if our remote gate were situated in a larger quantum circuit, it may be followed by a non-Clifford gate, which would necessitate the physical application of the Pauli frame at that point in time and we wish to account for this unfavourable scenario. Entanglement distillation has further variance in the latency because rounds of entanglement distillation fail probabilistically.

When using QED with post-selection, only some of the experimental shots are relevant, but the time taken by failed shots also matters. To account for this, we explicitly compute the success probability of each successful shot prior to each of the stabiliser measurements and then multiply the results together to find the total success probability of the shot. Concretely, we compute
\begin{equation} \label{eq:combining_probabilities}
    p_{\mathrm{s}} = p(X) p(Z|X),
\end{equation}
where $p(X)$ is the probability of a successful X stabiliser measurement and $p(Z|X)$ is the probability of a successful Z stabiliser measurement given that the preceding X stabiliser measurement was successful. Each of these probabilities is computed directly from the density matrix, $\rho$, immediately prior to the final Z-basis measurement used in the overall stabiliser measurement, using the expression
\begin{equation} \label{eq:success_prob_projection}
    p = \mathrm{tr}[(\1 \otimes \bra{0}) \rho (\1 \otimes \ket{0})],
\end{equation}
where $p=p(X)$ or $p(Z|X)$ and $\1 \otimes \ket{0}$ applies $\ket{0}$ to the ancilla qubit upon which the final Z-basis measurement is done. Equation \eqref{eq:success_prob_projection} computes the trace of the projection onto the successfully post-selected state.

We then divide the latency of the shot by the success probability to get the total latency and average this total latency over multiple successful shots to account for fluctuations in latency caused by different shots.

\section{Numerical results and discussion} \label{sec:results}

We evaluate the performance of the schemes introduced in Sec. \ref{sec:system_description} using a classical, event-based simulator that we built in Python 3.9 using the NetSquid library \cite{netsquid} and the related dqc\_simulator package that we wrote \cite{MySimGithub}, see Sec. \ref{sec:discrete_event_simulation}.  The primary figures of merit considered are the average gate fidelity, $\bar{F}_{\mathrm{out}}$, discussed in Sec. \ref{sec:average_gate_fidelity}, and the latency of the remote gate, discussed in Sec. \ref{sec:latency}. The results for the two variants of $\bar{F}_{\mathrm{out}}$ we consider, $\bar{F}_{\mathrm{MC}}$ and $\bar{F}_{\mathrm{A}}$, are discussed in Secs. \ref{sec:F_MC_Results} and \ref{sec:F_A_results}, respectively. The latency results are shown in Secs. \ref{sec:latency_results}. Finally, some further considerations that should be made when differentiating between the schemes are discussed in Sec. \ref{sec:further_considerations}.

\subsection{$\bar{F}_{\mathrm{MC}}$ results}\label{sec:F_MC_Results}

In order to identify the regimes in which the proposed ways of combatting noise are useful and find regimes in which different schemes do well, we consider multiple parameter regimes. As a starting point for our analysis, we use the hardware parameters for existing commercial devices. Where this is not possible, parameters from academic experiments are used. These state-of-the-art parameters are intended to approximate what could be achieved with current state-of-the-art technology, if that technology could be successfully integrated into a single device. All other parameter regimes are then based on improvements that could be made to the state of the art.

When choosing our state-of-the-art numbers, we use the hardware parameters from Quantinuum's H1 trapped-ion device \cite{quantinuum_specs}, where possible, because at the time the results were taken, this gave the best values that we could find for the parameters that we consider. Quantinuum's newer H2 and Helios devices have similar parameters, but with more qubits, and some architectural improvements, which was not relevant to our analysis. When values could not be readily found for the H1 device, numbers from IonQ Aria \cite{ionQAriaSpecs}, a competing commercial trapped-ion device, are used. Again, later generation devices exist but have similar parameter values for the parameters considered. Numbers related to the link between QPUs are sampled from those reported by Stephenson \textit{et al.} at the University of Oxford \cite{ionTrapEntDist94percent2m}. The relevant noise parameters are summarised in Table \ref{tab:noise_params}.%
\begingroup % for making setting only affect a localised area
\setlength{\tabcolsep}{10pt} % Column spacing, default value: 6pt
\renewcommand{\arraystretch}{1} % Row spacing, default value: 1
\begin{table}
    \centering
    \caption{The state-of-the-art values for the different parameters considered.}
    \begin{tabular}{ccc}\toprule
       Parameter & Value & Source  \\ \midrule
       $F_{\mathrm{w}}$ & $0.94$  & \cite{ionTrapEntDist94percent2m}  \\
       $\epsilon_{\mathrm{sg}}$ & $1.8 \times 10^{-5}$ &\cite{quantinuum_specs} \\
       $\epsilon_{\mathrm{tg}}$ & $9.7 \times 10^{-4}$ & \cite{quantinuum_specs} \\
       $\epsilon_{\mathrm{m}}$ & $2.33 \times 10^{-3}$ & \cite{quantinuum_specs} \\
       $r$ & $1$ Hz & \cite{ionQAriaSpecs}
    \end{tabular}
    \label{tab:noise_params}
\end{table}% 
\endgroup%

Similarly, the values we use for the durations of different operations, and their corresponding references, are listed in Table \ref{tab:durations}.%
\begingroup % for making setting only affect a localised area
\setlength{\tabcolsep}{10pt} % Column spacing, default value: 6pt
\renewcommand{\arraystretch}{1} % Row spacing, default value: 1
\begin{table}
    \centering
    \caption{The durations of all operations considered.}
    \begin{tabular}{ccc}\toprule
       Parameter & Value & Source  \\ \midrule
       Single-qubit gate time & $135 \, \mu$s  & \cite{ionQAriaSpecs}  \\
       Two-qubit gate time & $600 \, \mu$s & \cite{ionQAriaSpecs} \\
       Measurement time & 6 ms & Inferred from \cite{ionQAriaSpecs, AutoComm, metodi2005quantum} \\
       Classical comm. time & $10^{-8}$ s & \cite{ionTrapEntDist94percent2m, Paschotta_2006_fibers} \\
       Ebit. distr. time & $\frac{1}{182}$ s & \cite{ionTrapEntDist94percent2m}\\ \bottomrule
    \end{tabular}
    \label{tab:durations}
\end{table}% 
\endgroup%
The single-qubit gate time and two-qubit gate time used are sampled from the specifications for IonQ Aria \cite{ionQAriaSpecs}. The measurement time is estimated from the two-qubit gate time, based on the assertion made in Refs. \cite{AutoComm, metodi2005quantum} that the measurement time is five to ten times larger than the two-qubit gate time for trapped-ion quantum computers. We therefore choose the measurement time to be ten times the two-qubit gate time used. The classical communication time is taken to be the time to cross an optical fibre at $2 \times 10^{8}\,\mathrm{ms^{-1}}$ \cite{Paschotta_2006_fibers}. This neglects any classical processing times but gives a lower bound on the time expected. Actual classical communication times will most likely be larger than what we assume, but classical communication time is likely to be significantly lower than the latency of quantum operations and so should contribute comparatively negligibly to the overall latency and resultant memory depolarisation. For the ebit distribution time, $t_{\mathrm{dist}}$, we use $t_{\mathrm{dist}}=\frac{1}{R_{\mathrm{dist}}}$, where $R_{\mathrm{dist}} = 182$ Hz is the average entanglement distillation rate found in Ref. \cite{ionTrapEntDist94percent2m} between trapped ion systems $2 \, \mathrm{m}$ apart. Using an average time for entanglement distribution in this way is an approximation to reality which can have limitations when evaluating the performance of entanglement distillation \cite{zang_no-go_2025, zang_entanglement_2025}. In reality, entanglement distribution requires multiple attempts and success upon an earlier attempt will mean that an ebit spends more time depolarising on the memory than if it succeeds later. We upper bound the effect this will have by distributing entanglement sequentially for the entanglement distillation and QEC schemes that we consider.

\begin{figure}
    \centering
    \subfloat[]{\includegraphics[scale=0.6]{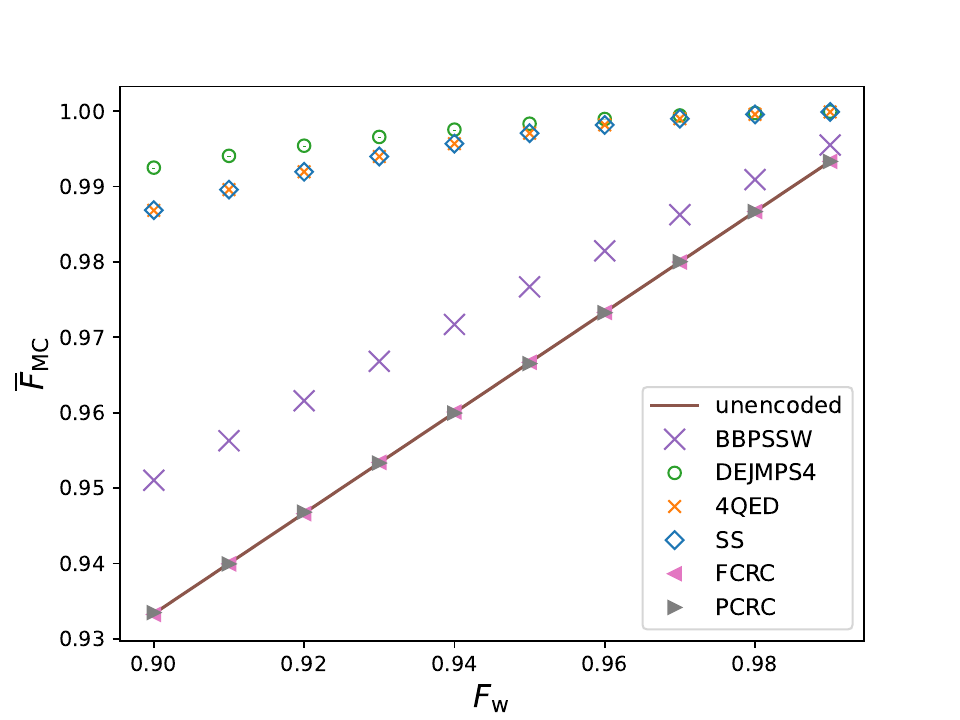}}
    \subfloat[]{\includegraphics[scale=0.6]{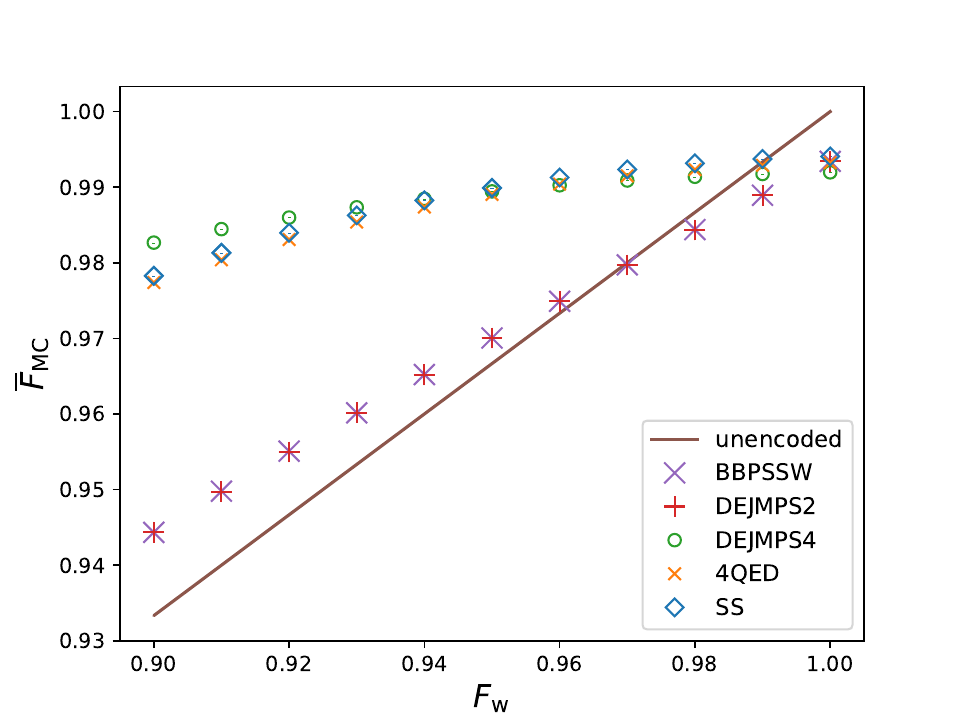}}    \caption{$\overline{F}_{\mathrm{MC}} $, averaged over all possible pure, separable input states to a remote CNOT gate as a function of $F_{\mathrm{w}}$ for (a) $\epsilon_{\mathrm{sg}} =  \epsilon_{\mathrm{tg}} = \epsilon_{\mathrm{m}} = r=0$, (b) $\epsilon_{\mathrm{sg}} =   1.8 \times 10^{-5}, \,  \epsilon_{\mathrm{tg}} = 9.7 \times 10^{-4}, \,   \epsilon_{\mathrm{m}} = 2.33 \times 10^{-3}  \text{ and } r=0.055\mathrm{Hz}$. Each data point is averaged over 40,000 $F_{\mathrm{out}}$ values generated with different input states uniformly sampled from the Bloch sphere of the control and target qubits. The different curves represent the results for the different error mitigation schemes considered in Secs. \ref{sec:qed_schemes} and \ref{sec:ent_dist_schemes}. As discussed in the main text, BBPSSW and DEJMPS are the entanglement distribution schemes from Refs. \cite{BBPSSW} and \cite{DEJMPS}, respectively; DEJMPS2 is two-ebit DEJMPS; DEJMPS4 is four-ebit DEJMPS; 4QED and SS are the LNCY \cite{412_subcode} code implemented using Fig. \ref{fig:encoding_circuits}(b) and Fig. \ref{fig:encoding_circuits}(c), respectively. Error bars show the standard error, which is too small to be visible here.}
    \label{fig:aoi}
\end{figure}

As is shown later, in Sec. \ref{sec:F_A_results}, current state-of-the-art devices have too much local noise for the proposed schemes to be beneficial relative to the encoded case and so we reduce the error, using the state-of-the-art parameters as starting points for finding regimes in which the results are useful. As a first step towards this, we obtain an upper bound on performance, by considering how each scheme performs in the presence of entanglement errors only. Given that the state-of-the-art value of $F_{\mathrm{w}} =0.94$, we consider the parameters $F_{\mathrm{w}} \in [0.9, 1], \, \epsilon_{\mathrm{sg}} = \epsilon_{\mathrm{tg}} = \epsilon_{\mathrm{m}} = r=0, \, $ and plot $\overline{F}_{\mathrm{MC}}$ as a function of $F_{\mathrm{w}}$. Each data point is taken by averaging over 40,000 points. The results of this analysis are shown in Fig. \ref{fig:aoi}(a). %
From Fig. \ref{fig:aoi}(a), the following observations can be made:
\begin{enumerate}
    \item The repetition code schemes, FCRC and PCRC, are indistinguishable from each other in the presence of entanglement error. It is interesting that there is no intrinsic entanglement error reduction benefit to using the more complicated FCRC scheme over the simpler PCRC scheme in our scenario. The PC-1TP method, shown in Fig. \ref{fig:qed_schemes}, benefits from a reduced gate count, greater simplicity and easier generalisation to other QED codes. For this reason, we use this scheme for the remaining codes investigated.
    \item Even in the absence of local errors, neither FCRC or PCRC shows any benefit, on average, over the unencoded remote gate. This result is unsurprising given that we consider depolarising noise, containing both bit and phase flip errors in equal measure, while the repetition codes that we consider only correct bit flip noise. Nonetheless, given that the repetition code demonstrates good performance in certain quantum communications protocols \cite{JingAlsina2020, rey-dominguez_rethinking_2025} and requires fewer physical qubits than alternative codes, it is valuable to explicitly verify that this is not the case here. From inspection of the data, significant advantage is found over the unencoded case for certain input states but for others they actually do worse than the unencoded case. This is mainly due to the aforementioned fact that the repetition code considered here does not detect the phase flip errors introduced by the depolarising noise that we consider. More details on the input state dependence of the repetition code performance are provided in Appendix \ref{app:repetition_code_failure}. However, we have been unable to infer a clear pattern as to which input states do better than others. 
    \item Entanglement distillation and both four-qubit error detection schemes, 4QED and SS, all provide a clear advantage over the unencoded case. The greatest advantage is provided by DEJMPS4 followed by 4QED and SS, with the least advantage offered by BBPSSW. It can be verified analytically that DEJMPS2 yields an identical entanglement fidelity to BBPSSW in this scenario and so is not shown. 
    \item 4QED and SS give essentially identical results. The difference between any two results always falls within the standard error range of the results. This is to be expected, as in the absence of local errors, there is very little to distinguish the encoders and decoders of the two schemes.
\end{enumerate}

We also consider the more realistic setting where finite local errors are present in the circuit. We use the parameters $F_{\mathrm{w}} \in [0.9, 1], \epsilon_{\mathrm{sg}} =   1.8 \times 10^{-5}, \,  \epsilon_{\mathrm{tg}} = 9.7 \times 10^{-4}, \,   \epsilon_{\mathrm{m}} = 2.33 \times 10^{-3}  \text{ and } r=0.055 \, \mathrm{Hz}$. The only difference compared to the state-of-the-art parameters is that $r$ has been reduced. Again, $\overline{F}_{\mathrm{out}}$ is plotted for $F_{\mathrm{w}}$ in the previously stated range to produce Fig. \ref{fig:aoi}(b). Based on Fig. \ref{fig:aoi}(b), we observe:
\begin{enumerate}
    \item As was the case previously, both of the four-qubit QED schemes considered, 4QED and SS, outperform both two-ebit entanglement distillation schemes. DEJMPS4, however, does a little better than 4QED and SS for $F_{\mathrm{w}} < 0.95$ and very slightly worse thereafter. The discrepancy between DEJMPS4 and 4QED/SS is always small for all considered $F_{\mathrm{w}}$ values.
    \item 4QED, SS and DEJMPS4 have a very clear advantage over the unencoded case until around $F_{\mathrm{w}} = 0.99$, at which point they perform similarly to the unencoded case. Therefore, there is a clear benefit to their use for this parameter regime. 
    \item DEJMPS2 and BBPSSW are almost identical, with at most $~0.01\%$ between them and neither scheme being consistently better for all $F_{\mathrm{w}}$.
    \item BBPSSW and DEJMPS2 outperform the unencoded case for $F_{\mathrm{w}} \lessapprox 0.97$ but offer a much smaller advantage than the DEJMPS4, 4QED and SS.
    \item SS slightly outperforms 4QED in the presence of the local error values considered, but the discrepancy between the two schemes remains very low. Overall, it seems that the slightly less gate efficient encoder and decoder used in 4QED makes very little difference in this scenario because the magnitude of the local errors are too low for a few gates to have a significant impact.
\end{enumerate}

Finally, we make use of the explicit averaging over input offered by Monte Carlo methods to consider the role played by input state. We again consider the parameters $\epsilon_{\mathrm{sg}} =  \epsilon_{\mathrm{tg}} = \epsilon_{\mathrm{m}} = r=0$ and $ \epsilon_{\mathrm{sg}} =   1.8 \times 10^{-5}, \,  \epsilon_{\mathrm{tg}} = 9.7 \times 10^{-4}, \,   \epsilon_{\mathrm{m}} = 2.33 \times 10^{-3}  \text{ and } r=0.055 \, \mathrm{Hz}$, which we call parameter set 1 and 2, respectively, but now only consider $F_{\mathrm{w}} \in [0.9, 0.99]$ as parameter set 1 would trivially give $\bar{F}_{\mathrm{out}} =1$ for $F_{\mathrm{w}}$ =1. The results are shown in Table \ref{tab:input_state}.

\begingroup % for making setting only affect a localised area
\setlength{\tabcolsep}{10pt} % Column spacing, default value: 6pt
\renewcommand{\arraystretch}{1} % Row spacing, default value: 1
\begin{table}
    \centering
    \caption{The percentage difference between the minimum and maximum values of the output fidelity, $\Delta_{\mathrm{mm}} = \frac{(F_{\mathrm{out}})_{\mathrm{max}} - (F_{\mathrm{out}})_{\mathrm{min}}}{(F_{\mathrm{out}})_{\mathrm{min}}} \times 100$, where $(F_{\mathrm{out}})_{\mathrm{max}}$ and $(F_{\mathrm{out}})_{\mathrm{min}}$ are the maximum and minimum values of $F_{\mathrm{out}}$, respectively, for a given $F_{\mathrm{w}}$, and all noise parameters are fixed, so that only the input state of the control and target qubits is varied. The minimum and maximum values of $\Delta_{\mathrm{mm}}$ shown are obtained when $F_{\mathrm{w}} = 1$ and $F_{\mathrm{w}} = 0.9$, respectively.}
    \begin{tabular}{cccc}\toprule
       Scheme & Parameter set & Min. $\Delta_{\mathrm{mm}}$ (\%) & Max. $\Delta_{\mathrm{mm}}$ (\%) \\ \midrule
       BBPSSW &  1 &  0.671 & 7.10 \\ 
       BBPSSW & 2 & 1.02 & 7.09 \\
       DEJMPS2 & 2 & 1.09 & 7.53 \\
       DEJMPS4 & 1 & $4.51 \times 10^{-3}$ & 0.507 \\
       DEJMPS4 & 2 & 0.71 & 2.12 \\ 
       FCRC & 1 & 1.01 & 11.0 \\
       PCRC & 1 & 1.01 & 11.0 \\
       4QED & 1 & $4.51 \times 10^{-3}$ & 0.504 \\
       4QED & 2 & 0.277 & 0.886 \\
       SS & 1 & $4.51 \times 10^{-3}$ & 0.504 \\
       SS & 2 & 0.121 & 0.737 \\
    \bottomrule
    \end{tabular}
    \label{tab:input_state}
\end{table}% 
\endgroup%

From Table \ref{tab:input_state}, we can see that altering the input state can produce percentage differences as large as $\Delta_{\mathrm{mm}} = 11.0$ \% for some error handling schemes when $F_{\mathrm{w}} = 0.9$ \% and the percentage difference can still reach up to $\Delta_{\mathrm{mm}} = 1.09\%$ when $F_{\mathrm{w}} = 0.99$ for DEJMPS2. As such, it is clearly important to account for the role played by the input state, especially when the entanglement error is high. This can be done explicitly, as we do in this section, or implicitly, as is done in the next section, using the methods discussed in Sec. \ref{sec:average_gate_fidelity}.

\subsection{$\bar{F}_{\mathrm{A}}$ results} \label{sec:F_A_results}

To generalise the results from the previous section to arbitrary input states, we now consider $\bar{F}_{\mathrm{out}} = \bar{F}_{\mathrm{A}}$. The results for each data point are averaged over 50 runs, to account for fluctuations in the latency, and therefore memory depolarisation applied, which occur due to the measurement-dependent aspects of the circuit. Having observed in Sec. \ref{sec:F_MC_Results} that BBPSSW and DEJMPS2 perform very similarly and that this is also true of 4QED and SS, we now consider a restricted set of schemes: unencoded, DEJMPS2, DEJMPS4 and SS.

\begin{figure}
    \centering
    \includegraphics[width=0.5\linewidth]{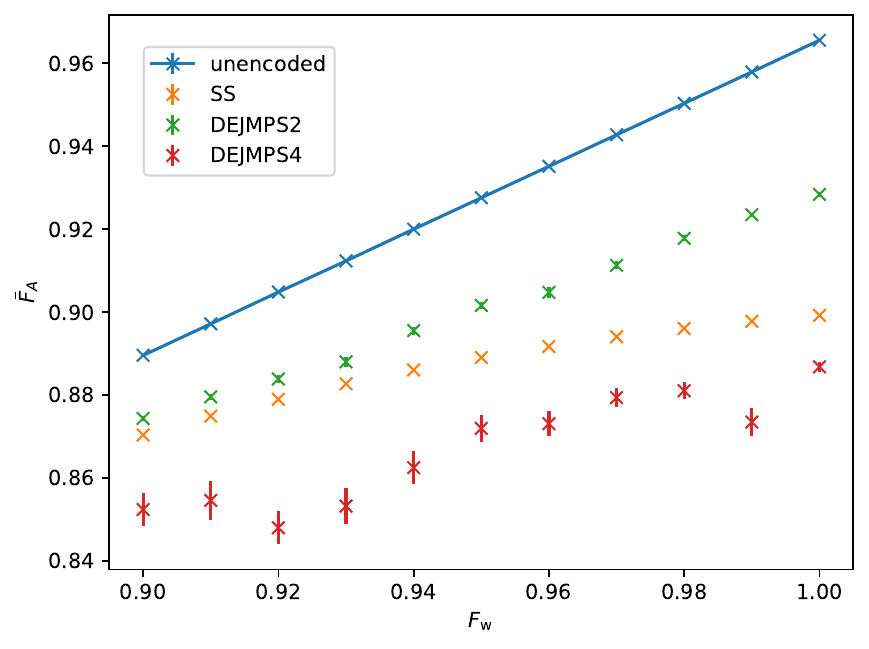}
    \caption{$\bar{F}_{\mathrm{A}}$ for parameters $F_{\mathrm{w}} \in [0.9, 1], \epsilon_{\mathrm{sg}} =   1.8 \times 10^{-5}, \,  \epsilon_{\mathrm{tg}} = 9.7 \times 10^{-4}, \,   \epsilon_{\mathrm{m}} = 2.33 \times 10^{-3}  \text{ and } r=1 \, \mathrm{Hz}$. All data points are averaged over 50 runs and error bars indicate the standard error.}
    \label{fig:F_A_soa}
\end{figure}

We first verify the claim, made in Sec. \ref{sec:F_MC_Results}, that no benefit is seen for the error reducing schemes over the unencoded case, when the state-of-the-art parameters in Table \ref{tab:noise_params} are used. The results can be seen in Fig. \ref{fig:F_A_soa}, which shows  $\bar{F}_{\mathrm{A}}$ for the parameters $F_{\mathrm{w}} \in [0.9, 1], \epsilon_{\mathrm{sg}} =   1.8 \times 10^{-5}, \,  \epsilon_{\mathrm{tg}} = 9.7 \times 10^{-4}, \,   \epsilon_{\mathrm{m}} = 2.33 \times 10^{-3}  \text{ and } r=1 \, \mathrm{Hz}$. We note that, when producing the figure, we removed one erroneous value for which the fidelity was of the order of $10^3$. We believe that this was due to pathological floating point errors when computing the fidelities of the output density matrices, as in two other cases the SciPy library underlying parts of NetSquid raised an error in the fidelity function used. We also found, when considering different parameter regimes not reported here, to search for such errors, we would occasionally obtain an unphysical result during the fidelity calculations. We could reproduce a physically reasonable fidelity value when calculating fidelities using density matrices manually produced from printed, and therefore truncated, representations of the actual density matrices, again implying that the issue was a numerical error during the fidelity calculation itself. %

From Fig. \ref{fig:F_A_soa}, it is clear that DEJMPS2/4 and SS perform worse than the unencoded case with all considered data points falling well below those for the unencoded case. Importantly, this is true even when there is no entanglement error, indicating that local errors, in of themselves, render the entanglement distillation and QED schemes considered unviable. It is also noteworthy that the relative performance of the schemes is very different compared to the results with a lower memory depolarisation rate shown in Fig. \ref{fig:aoi}, with DEJMPS2 outperforming both SS and DEJMPS4 despite using fewer resources. These observations imply that increasing the number of rounds of DEJMPS further is unlikely to help and that improved experimental parameters are likely to be needed for the practical implementation of a remote CNOT gate. 

\begin{figure}
    \centering
    \subfloat[]{\includegraphics[scale=0.6]{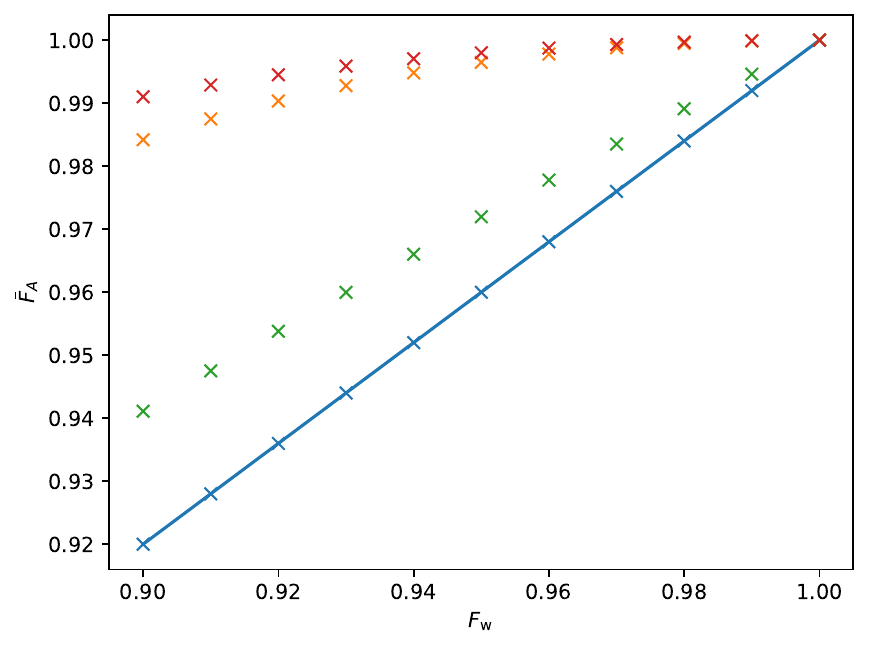}}
    \subfloat[]{\includegraphics[scale=0.6]{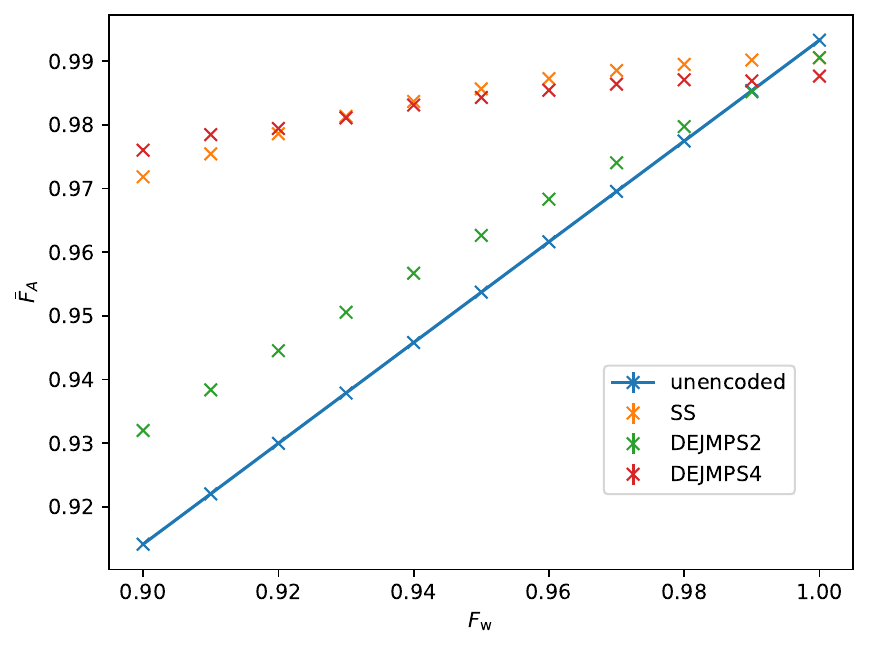}}
    \caption{$\bar{F}_{\mathrm{A}}$ as a function of $F_{\mathrm{w}}$ for (a) $\epsilon_{\mathrm{sg}} =  \epsilon_{\mathrm{tg}} = \epsilon_{\mathrm{m}} = r=0$, (b) $\epsilon_{\mathrm{sg}} =   1.8 \times 10^{-5}, \,  \epsilon_{\mathrm{tg}} = 9.7 \times 10^{-4}, \,   \epsilon_{\mathrm{m}} = 2.33 \times 10^{-3}  \text{ and } r=0.055\mathrm{Hz}$.  All data points are averaged over 50 runs and error bars indicate the standard error, which is too small to be visible here.}
    \label{fig:reproducing_Fout}
\end{figure}

Having verified the claim that state-of-the-art parameters are insufficient, we then consider the same experimental parameters as in Sec. \ref{sec:F_MC_Results}, resulting in Figs. \ref{fig:reproducing_Fout}(a) and \ref{fig:reproducing_Fout}(b) for parameter regimes $\epsilon_{\mathrm{sg}} =  \epsilon_{\mathrm{tg}} = \epsilon_{\mathrm{m}} = r=0$ and $F_{\mathrm{w}} \in [0.9, 1], \, \epsilon_{\mathrm{sg}} =   1.8 \times 10^{-5}, \,  \epsilon_{\mathrm{tg}} = 9.7 \times 10^{-4}, \,   \epsilon_{\mathrm{m}} = 2.33 \times 10^{-3}, \, r=0.055 \, \mathrm{Hz}$, respectively.

The results shown in Fig. \ref{fig:reproducing_Fout} for $\bar{F}_{\mathrm{A}}$ are broadly similar to those for $\bar{F}_{\mathrm{MC}}$ shown in Fig. \ref{fig:aoi}, indicating that arbitrary inputs and those of pure separable input states perform similarly here. Again, we see that SS and DEJMPS4 perform very similarly and significantly outperform DEJMPS2, indicating that if the memory depolarisation rate is low then it is worthwhile to expend more resources in combatting errors. All schemes outperform the unencoded case, showing that even the quite small QED and entanglement distillation schemes considered here can yield significant improvements over the unencoded case, if the memory depolarisation rate is sufficiently low.

Comparing Figs. \ref{fig:F_A_soa} and Fig. \ref{fig:reproducing_Fout}(b), we see that there must exist a transition point between higher values of $r$ for which SS performs better than DEJMPS4 and smaller values of $r$ for which DEJMPS4 performs better. In the future, it would be interesting to explore if there is a regime in which SS does better than the unencoded case and both entanglement distillation schemes.

\subsection{Latency results} \label{sec:latency_results}

Considering the $\overline{F}_{\mathrm{MC}}$ and $\bar{F}_{\mathrm{A}}$ results alone, there is little to distinguish between SS and DEJMPS4. However, it does not matter what the quality of the output from the remote gate is if the time taken to carry it out is prohibitive. As such, we consider the latency of implementing the remote gate using the different schemes here. 

\begin{figure}
    \centering
    \includegraphics[width=0.5\linewidth]{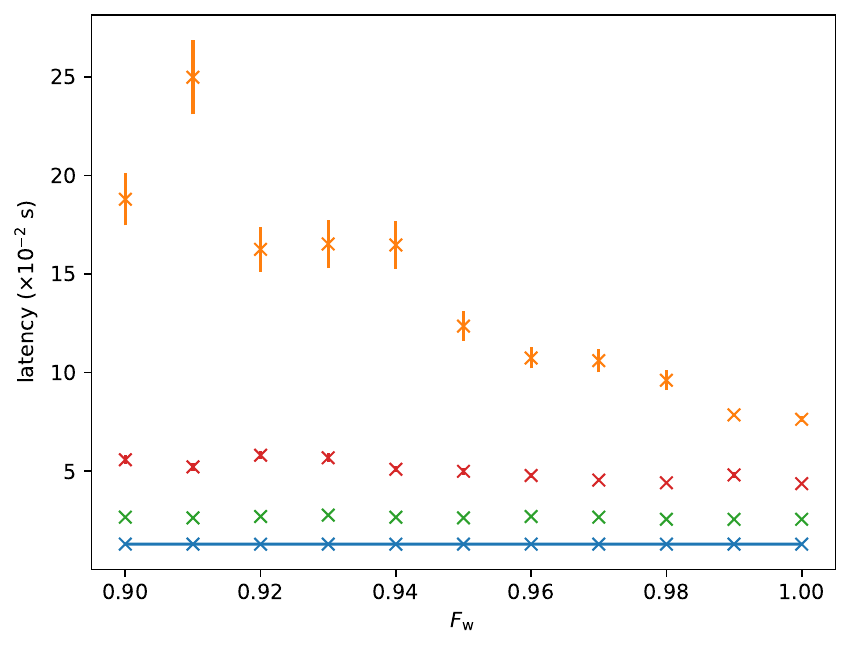}
    \raisebox{2cm}{\includegraphics[scale=0.1]{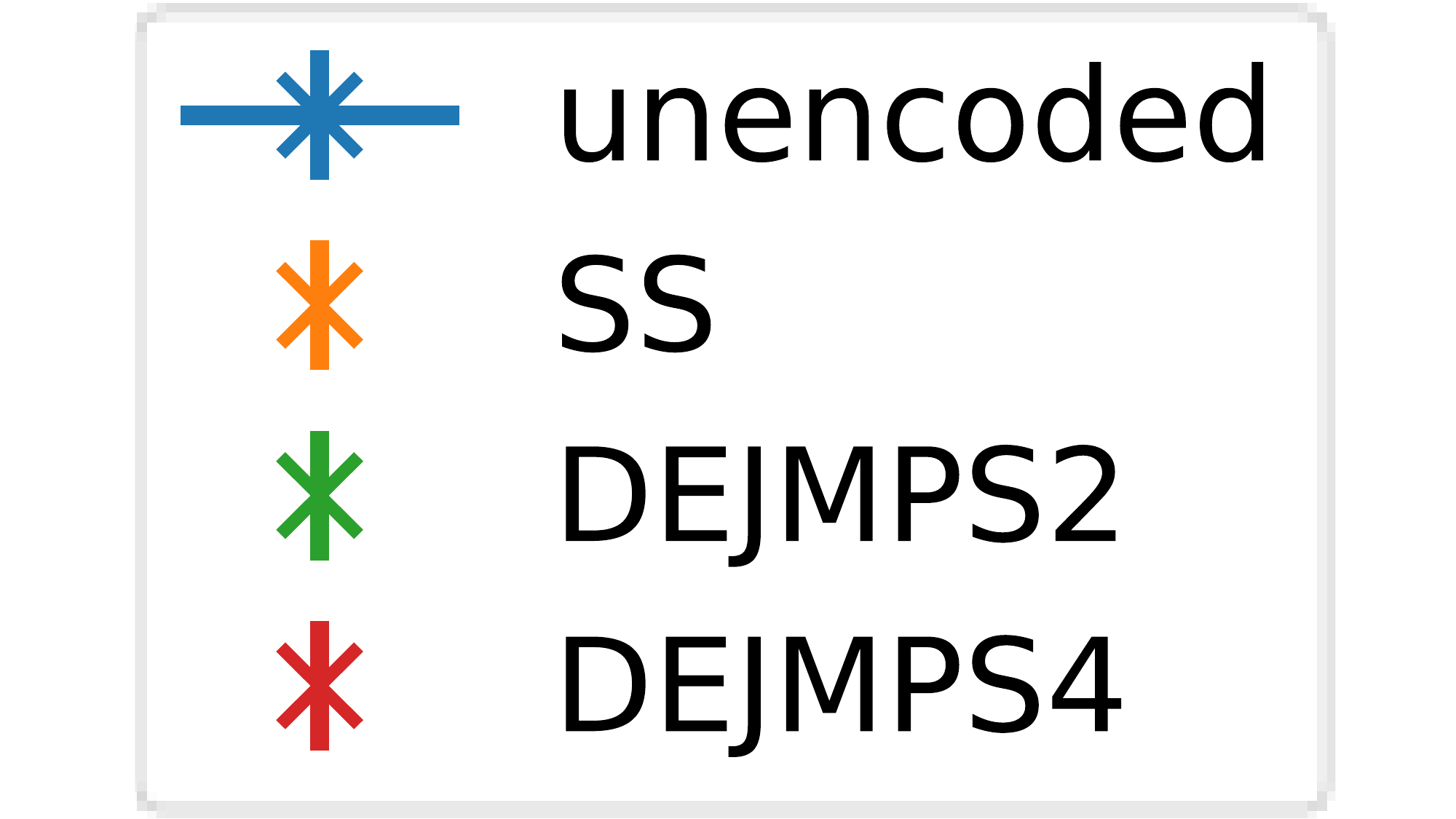}}
    \caption{The latency for $\epsilon_{\mathrm{sg}} =   1.8 \times 10^{-5}, \,  \epsilon_{\mathrm{tg}} = 9.7 \times 10^{-4}, \,   \epsilon_{\mathrm{m}} = 2.33 \times 10^{-3}  \text{ and } r=0.055\mathrm{Hz}$.  All data points are averaged over 50 runs and error bars indicate the standard error.}
    \label{fig:latency}
\end{figure}

Figure \ref{fig:latency} shows the latency of the different schemes averaged over 50 runs of the discrete-event simulation for the parameters $F_{\mathrm{w}} \in [0.9, 1], \epsilon_{\mathrm{sg}} =   1.8 \times 10^{-5}, \,  \epsilon_{\mathrm{tg}} = 9.7 \times 10^{-4}, \, \epsilon_{\mathrm{m}} = 2.33 \times 10^{-3}, \, \text{ and } r=0.055 \, \mathrm{Hz}$. From the figure, we observe that SS has the highest latency, followed by DEJMPS4 and then DEJMPS2. SS also seems to be much more sensitive to increases in ebit error than entanglement distribution. However, it is important to highlight the fact that we consider the ebit-constrained regime in which the teleportation of the qubits in SS are done sequentially and the entanglement distributions in the entanglement distillations are also performed sequentially. This is not optimal and is intended to lower bound the performance and limit the usage of communication qubits. If communication qubits are plentiful and parallelisation could be maximised then we would expect SS to benefit the most from this in terms of latency because all of its teleportations could be done in a single round, whereas DEJMPS is an iterative process and so inherently requires more sequential entanglement distributions. This could also have a meaningful impact on $\bar{F}_{\mathrm{MC}}$ and $\bar{F}_{\mathrm{A}}$, if the memory depolarisation rate is sufficiently high. Due to limitations of the scheduler for the dqc\_simulator package we use to perform the teleportations in SS and the additional computational cost of working with the much larger density matrices encountered in the maximally parallelised regime, we defer explorations of this regime to future work.

That said, when discussing the latency of an individual remote gate, it is important to contextualise the individual remote gate we consider in this work within the broader distributed quantum computing context in which multiple remote gates may be present. A key facet of this is the fact that failed rounds of DEJMPS only require communication qubits to be discarded and so do not destroy any quantum information required to complete the algorithm, which is stored in the processing qubits. The ebits used in DEJMPS are a consumable resource and the only cost to discarding them is time. By contrast, if SS fails then the entire algorithm must be restarted as there is a detected error in the processing qubits. This means that failure is much more costly for QED than entanglement distillation. This tells us that when considering distributed quantum circuits in the limit where errors in ebit errors are likely, entanglement distillation will have better latency performance than SS. The point at which such a limit will occur is dictated by the error probability of each remote gate and the number of remote gates.

\subsection{Further considerations}\label{sec:further_considerations}

In addition to $\bar{F}_{\mathrm{out}}$ and latency, there are other practical considerations, when situating remote gates within the wider distributed quantum computing context. The first of these is that SS requires more qubits to implement than DEJMPS4, as SS requires an additional three processing qubits on one QPU, with which to encode the control qubit prior to teleportation. 

Another area in which there is some distinction is the number of communication qubits each scheme requires to be available in parallel. As noted in Sec. \ref{sec:unencoded_case}, it is possible that communication qubits could be a different atomic or ionic species to the processing qubits, or, for other qubit modalities, equivalently different in some way. As such, they may need different control systems. They will also require additional optical apparatus to entangle them with photons and facilitate ebit distribution. For example, communication qubits may require proximity to an optical lens for coupling the photons they emit to a fibre or may be in a cavity. For these reasons, it is plausible that the presence of communication qubits on a QPU may degrade the performance of the processing qubits. Unwanted coupling of processing qubits to optical components or a higher cost associated with gates or unwanted interactions between different ionic species. Moreover, communication qubits may also impose engineering constraints, due to the need for additional control apparatus. In such scenarios, it may be desirable to minimise the number of communication qubits required, even if this decreases the parallelisation of ebit distribution possible and, in doing so, increases the latency and memory depolarisation. Similar constraints may also arise due to limitations in network connectivity or routing challenges, as it may be challenging to connect arbitrarily many communication qubits to arbitarily many communication qubits in another QPU.

The minimum number of communication qubits per QPU required by different noise mitigation schemes are summarised in Table \ref{tab:comm_qubits}, along with the number of communication qubits required to maximise parallelisation and minimise latency. For SS, the teleportation of each of the physical qubits, encoding the logical state to be teleported, can be done sequentially, and so only one communication qubit is needed per QPU, assuming that the teleported state is swapped into a processing qubit after each teleportation. By contrast, the entanglement distillation schemes that we consider need to use at least two communication qubits on each QPU per round of distillation and at least one of those must be retained for the next round. Concretely, in DEJMPS4, we would minimise the number of communication qubits used by splitting the first round into two sequential steps. In the first step, two ebits are distilled into one higher quality ebit, as shown in Fig. \ref{fig:entanglement_distillation_QC}(b). This requires two communication qubits on each QPU. One of these is measured out as part of the distillation process and can then be re-initialised and re-used in the second step. In the second step, an ebit is distributed between the re-initialised communication qubits on both QPUs, and another ebit is distributed using a fresh communication qubit on each QPU. These are again distilled as shown in Fig. \ref{fig:entanglement_distillation_QC}(b). Meanwhile, a third communication qubit on each QPU is being used to retain the distilled ebit from step 1. These two distilled pairs are then distilled again to produce a higher quality pair. As such, there must be at least three qubits available to do DEJMPS4 and this number holds constant for an arbitrary number of rounds of DEJMPS because we can always distil one pair over several rounds and then retain it in the memory while distilling another pair over several rounds. It should be noted however that doing so is not guaranteed to improve the fidelity of the distilled ebits due to memory depolarisation or memory decoherence while the first distilled ebit produced waits for distillation in the other ebit to occur.
\begingroup % for making setting only affect a localised area
\setlength{\tabcolsep}{10pt} % Column spacing, default value: 6pt
\renewcommand{\arraystretch}{1} % Row spacing, default value: 1
\begin{table}
    \centering
    \caption{The minimum and maximum numbers of communication (comm.) qubits per QPU required by different noise mitigation schemes.}
    \begin{tabular}{ccc}\toprule
       Scheme & Min. \# comm. qubits & \# comm. qubits for max. parallelisation \\ \midrule
      DEJMPS2/BBPSSW & 2 & 2  \\ 
      DEJMPS4  & 3  & 4 \\
      SS & 1 & 4 \\
    \bottomrule
    \end{tabular}
    \label{tab:comm_qubits}
\end{table}% 
\endgroup%

As well as flexibility in the number of communication qubits required, QED also may also be able to make different noise-reduction trade-offs compared to entanglement distillation. If the noise parameters are sufficiently low, we can reduce the error in our ebits by increasing the distance of the code, at the cost of requiring more physical qubits. If we have sufficient communication qubits such that entanglement distribution can be done in parallel then the latency of a single PC-1TP gate scales only with the size of the encoding and stabiliser circuits and is independent of the number of entanglement distributions required. By contrast, entanglement distillation requires sequential rounds to improve the fidelity. As such, in regimes where entanglement distribution has a comparatively high latency relative to local gates, QED-based schemes may be able to offer better time-based trade-offs for noise than entanglement distillation, although this is highly contingent on the speed with which the encoding and decoding circuits can be done.

\section{Conclusion} \label{sec:conclusion}

In this work, we applied quantum error detection to remote gates as a means of improving fidelity and investigated the performance of those schemes and entanglement distillation in increasing the fidelity of remote gates within the quantum data centre context. Such remote gates form the fundamental building blocks of gate-based distributed quantum computing. 

We focus on small schemes with relatively low qubit overheads and consider the regime in which quantum communication is constrained and so entanglement distribution is performed sequentially, which provides a lower bound on the expected performance. In practice, such scenarios may occur due to limited numbers of communication qubits or challenges with the scheduling constraints. 

With these restrictions, we find that state-of-the-art experimental parameters are not sufficient for the small schemes we consider to outperform an unencoded remote gate. However, if the memory depolarisation rate, or equivalently the latency of the remote gates, can be sufficiently reduced then the schemes considered are able to significantly outperform the unencoded case, even with noisy gates and measurements.

Under a low memory depolarisation regime, with sequential entanglement distributions, our results suggest that, of the error reduction schemes considered, entanglement distillation performs the best, as it is able to improve the average output fidelity better or comparably to the error detection schemes investigated, with fewer qubits and a lower latency cost. Both entanglement distillation schemes are potentially useful and there is a trade-off between the better fidelity improvement of four-ebit, two-round DEJMPS entanglement distillation and the faster implementation and lower qubit cost of two-ebit, one-round DEJMPS. However, the significant improvement in output fidelity of four-ebit DEJMPS relative to two-ebit DEJMPS most likely makes it worth doing at least two rounds of entanglement distillation.

On the other hand, we expect that QED may perform better in a maximally parallelised regime, where all entanglement distributions can be done in parallel, if the latency cost of entanglement distribution is much higher than that of local gates. QED can support greater parallelisation of ebits than entanglement distillation, due to the inherently iterative nature of entanglement distillation, and consequently, QED may achieve a lower latency, and therefore lower memory depolarisation, than entanglement distillation, in such a regime. Moreover, QED may also make sense in scenarios where the number of communication qubits is very limited or there are challenges with connecting communication qubits between arbitary QPUs within a quantum data centre. It requires fewer communication qubits to be available in parallel to implement the partially coded QED-based scheme proposed than entanglement distillation. 

An alternative regime, not considered in this work, would be to perform error detection only on the ebit resources and decode prior to beginning teleportation, which may offer better performance. Future work on QED in the distributed quantum computing context may also be of theoretical interest, as a pathway towards fault-tolerant quantum error correction, which is harder to implement. It would also be interesting to investigate how well QED and entanglement distillation perform when used together.

Overall, this work indicates that both QED and entanglement distillation may be powerful tools for building near to mid term demonstrations of distributed quantum computers. We investigate and discuss a number of considerations that should be considered when making use of these tools and also motivate many areas for future exploration. 

\section{Acknowledgements}

We would like to acknowledge funding from the Leeds Doctoral Scholarship, the UK EPSRC grants EP/Y037421/1 and EP/X040518/1, and UKRI grant UKRI598. We would also like to thank Simon Milz, Jiannis K Pachos, and Elham Kashefi for useful discussions about this work.

\section{Data availability statement}

The data that support the findings of this study are openly available at the following URL/DOI: https://doi.org/10.5281/zenodo.21225458

\appendix

\section{QED as a quantum channel}\label{app:qed_channel}

In Sec. \ref{sec:average_gate_fidelity} of the main text, it is pointed out that the applicability of Eq. \eqref{eq:average_gate_entanglement_fidelity_relation} to a remote gate implemented using QED is non-trivial. Here, we provide some justification for our assumption that Eq. \eqref{eq:average_gate_entanglement_fidelity_relation} holds for our analysis of such scenarios.

To see why it is non-trivial to consider the applicability of Eq. \eqref{eq:average_gate_entanglement_fidelity_relation} to a FC-1TP or PC-1TP remote gate, consider the action of either of the stabiliser measurements used in the implementation of these schemes. The stabiliser measurement circuits are shown in Fig. \ref{fig:decoding_circuits} (b). We post-select states for which the final Z-basis measurement has the value 0. Denoting such an operation as $\mathcal{E}$, the normalised state produced by acting $\mathcal{E}$ on an input state, $\rho$, is \cite{NielsenChuang}
\begin{equation}
    \frac{\mathcal{E} (\rho)}{\mathrm{tr}[\mathcal{E}(\rho)]},
\end{equation}
where $\mathrm{tr}$ is the standard trace operation and $p_{\mathrm{s}}(\rho) = \mathrm{tr}[\mathcal{E}(\rho)]$ can be interpreted as the probability of successful post-selection \cite{NielsenChuang}. Clearly, if $p_{\mathrm{s}} \neq 1$ for all $\rho$ then  $\mathcal{E}$ is not trace-preserving, which is required for a quantum channel. Physically, the requirement of trace-preservation enforces the fact that $\mathcal{E}$ fully describes the quantum process being considered \cite{NielsenChuang}. However, we are only interested in the average gate fidelity of scenarios where post-selection is successful, for which we can define a normalised post-selection operation as $\Lambda(\rho) = \frac{\mathcal{E} (\rho)}{\mathrm{tr}[\mathcal{E}(\rho)]}$. 

On the other hand, $\Lambda$ is not necessarily a convex linear map on the set of density matrix, as required by a quantum channel \cite{NielsenChuang}. That is to say, 
\begin{equation}\label{eq:non_linearity}
    \begin{aligned}
    \Lambda(\sum_i p_i\rho_i) &= \frac{\mathcal{E} (\sum_i p_i \rho_i)}{\mathrm{tr}[\mathcal{E} (\sum_i p_i \rho_i)]} \\
    &= \frac{\sum_i p_i \mathcal{E} (\rho_i)}{\mathrm{tr}[\sum_i p_i \mathcal{E} (\rho_i)]} \\
    &\neq \sum_i p_i \Lambda(\rho_i),
    \end{aligned}
\end{equation}
in general, for probabilities $p_i$. Equality does not hold on the final line due to the non-linearity of the trace.

Despite this, if the success probability, $p_{\mathrm{s}} = \mathrm{Tr[\sum_i p_i \mathcal{E} (\rho_i)]}$, is constant with respect to the input state then the penultimate line of Eq. \eqref{eq:non_linearity} becomes
\begin{equation}
    \begin{aligned}
        \sum_i p_i \frac{\mathcal{E}(\rho_i)}{p_s} = \sum_i p_i \Lambda(\rho_i),
    \end{aligned}
\end{equation}
as required. Therefore, for  the normalised successful post-selection operation, $\Lambda$, that we consider, it suffices to verify that the success probability is constant with respect to the input state. We expect this should be the case, due to the assumption of independent errors that we have made in our modelling, which have no bearing on input state. We are not trying to distinguish between errors or do any complicated decoding and so there is not reason for the stabiliser measurements to detect errors on one input state over another. They are sensitive only to the number of errors in the circuit.  

To support this reasoning, we numerically simulate PC-1TP with the LNCY code. We consider 100 pure, two-qubit states randomly sampled using the two-qubit Haar measure and use Eqs. \eqref{eq:combining_probabilities} and \eqref{eq:success_prob_projection} to numerically compute the success probability for the noise parameters $F_{\mathrm{w}} \in [0.9, 1], \epsilon_{\mathrm{sg}} =   1.8 \times 10^{-5}, \,  \epsilon_{\mathrm{tg}} = 9.7 \times 10^{-4}, \,   \epsilon_{\mathrm{m}} = 2.33 \times 10^{-3}  \text{ and } r=1 \, \mathrm{Hz}$. We find that, for a given set of parameter values, the numerical values for the success probability agree up to 14 significant figures, but some disagreement between the computed values is observed in the 15th and 16th significant figures, most likely due to floating point error. We also use the same methodology to exhaustively check the success probability for all basis states in the two-qubit Pauli basis $\{\1, X, Y, Z\}^{\otimes 2}$ and the basis states created by multiplying both qubits by Hadarmards for every state in the Pauli basis. Additionally, to explicitly include entangled states, we  consider the Bell basis of maximally entangled states. Again, we find that the success probabilities agree up to 14 significant figures in all cases. In light of this very high degree of agreement between the wide variety of pure states that we consider, we conjecture that any pure states are likely to show similar levels of agreement. If this is the case, then we would also expect arbitrary mixed states to show the same level of agreement. To see this, consider an arbitrary state, $\rho$. As before, the success probability, $p_{\mathrm{s}}(\rho)$ is given by
\begin{equation}
        p_{\mathrm{s}} (\rho) = \mathrm{tr}[\mathcal{E}(\rho)].
\end{equation}
Representing the density matrix as $\rho = \sum_i p_i \rho_i$, where $\rho_i = \ket{i}\bra{i}$ is some pure state, we find
\begin{equation}
    \begin{aligned}
        p_{\mathrm{s}}(\rho) &= \mathrm{tr}[\mathcal{E}(\sum_i p_i \rho_i))] \\
        &= \sum_k \bra{k} \sum_i p_i \mathcal{E}(\rho_i) \ket{k} \\
        &= \sum_i p_i \sum_k \bra{k} \mathcal{E} (\rho_i) \ket{k} \\
        &= \sum_i p_i \mathrm{tr}[\mathcal{E} (\rho_i)].
    \end{aligned}
\end{equation}
If the success probability of each pure $\rho_i$, $\mathrm{tr}[\mathcal{E} (\rho_i)]$, is a constant, $C$, for all $i$ and $j$, which we have numerically verified holds up to 14 significant figures for the various input states we considered, then $p_{\mathrm{s}} (\rho)$ becomes
\begin{equation}
        C \underbrace{\sum_i p_i}_{=1} = C.
\end{equation}
We emphasise that in our discussion above, the value of $C$ is found numerically for a variety of input states rather than being an exact constant, although, for the input states considered, the agreement is very good, as discussed above. In light of this, we assume that the success probability is approximately constant for arbitrary input states. We also note that the value of $C$ is different for different noise parameters. In the absence of an analytical expression for $\Lambda$, we consider each individual set of noise parameters to yield a different quantum channel, whose average gate fidelity we then find. Variations in latency due to the probabilistic application of Pauli gates during teleportation also yield separate channels and so we fix the latency in the analysis discussed above by fixing the simulation seed.

As such, we assume that $\Lambda$ approximates a linear CPTP map, as required for Eq. \eqref{eq:average_gate_entanglement_fidelity_relation} to be applied to it. We have provided numerical evidence for its linearity; trace preservation holds by construction, as we explicitly normalise the state; and complete positivity follows immediately from the complete positivity of $\mathcal{E}$.

However, as a sanity check, providing further validation of the correctness of the procedure for a remote gate with QED, we evaluate some properties of the Choi matrix of our normalised post-selection operator, $\Lambda$, to confirm that $\Lambda$ approximates a CPTP map as demonstrated above. A general linear map, $\mathcal{E}$, with Choi matrix $\tilde{\mathcal{E}}$, is completely positive if and only if $\tilde{\mathcal{E}}$ is positive semi-definite and trace preserving if and only if $\mathrm{tr}_R(\tilde{\mathcal{E}}) = \frac{1}{d} \1$ \cite{ekert_qubitguide_2025}, where $\mathrm{tr}_R$ is the partial trace over the reference system, $R$, of the Choi matrix that we construct to take our average gate fidelity results. We simulate the Choi matrix as discussed in Sec. \ref{sec:average_gate_fidelity}, for the parameters $F_{\mathrm{w}} \in [0.9, 1], \epsilon_{\mathrm{sg}} =   1.8 \times 10^{-5}, \,  \epsilon_{\mathrm{tg}} = 9.7 \times 10^{-4}, \,   \epsilon_{\mathrm{m}} = 2.33 \times 10^{-3}  \text{ and } r=1 \, \mathrm{Hz}$, and find that $\tilde{\Lambda}$ is positive semidefinite and $\mathrm{tr}_R(\tilde{\Lambda}) = \frac{1}{4} \1$, which is consistent with the value of $d$ that we use ($d=4$). 

\section{Input states and the repetition code performance} \label{app:repetition_code_failure}

In the second observation on Fig. \ref{fig:aoi}(a) that we make in Sec. \ref{sec:results}, we note that neither FCRC nor PCRC shows any benefit on average over the unencoded case, but there is a significant advantage to using FCRC/PCRC for some input states and a disadvantage in other cases. Here, we expand upon that statement.

\begin{figure}
    \centering
    \includegraphics[scale=0.6]{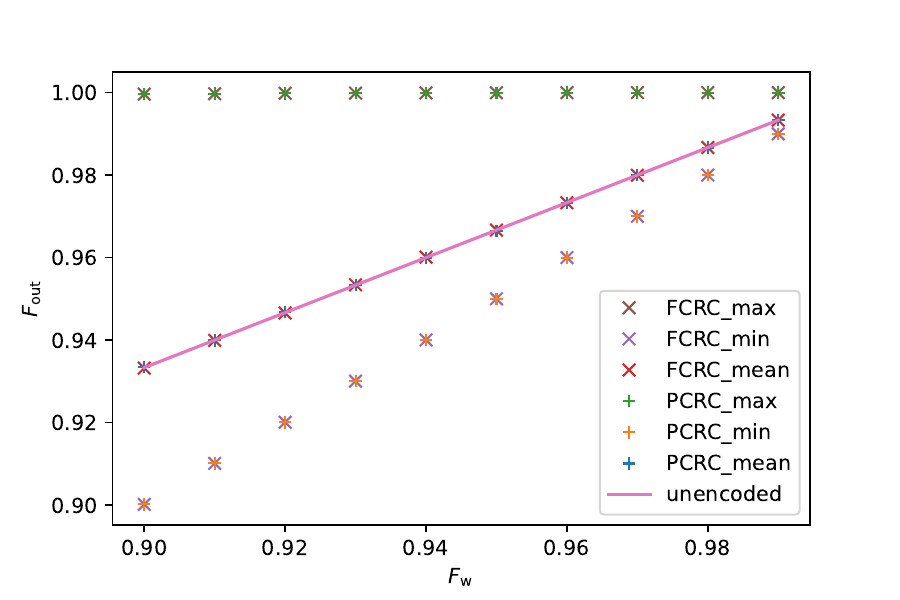}
    \caption{The min, max and mean $F_{\mathrm{out}}$ with respect to the parameters $F_{\mathrm{w}} \in [0.90, 0.99], \, \epsilon_{\mathrm{sg}} = \epsilon_{\mathrm{tg}} = \epsilon_{\mathrm{m}} = r=0,$ for a remote CNOT gate encoded using FCRC and PCRC.}
    \label{fig:understand_input_state}
\end{figure}

Figure \ref{fig:understand_input_state} %
shows the minimum, maximum and mean, values of $F_{\mathrm{out}}$ for a remote CNOT gate implemented using FCRC and PCRC. The parameters are $F_{\mathrm{w}} \in [0.90, 0.99], \, \epsilon_{\mathrm{sg}} = \epsilon_{\mathrm{tg}} = \epsilon_{\mathrm{m}} = r=0.$ For each datapoint, the result is sampled from 40,000 raw data points for which no error was detected. From Fig. \ref{fig:understand_input_state}, we can see that for some input states, both FCRC and PCRC correct almost all of the error. However, for other input states, the encoded case actually does worse than the unencoded case, even in the absence of local errors. In such scenarios, it seems that the additional phase errors imposed on the system by using additional imperfect ebits, counteracts any benefit to the repetition code error detection, which detects only bit flips. These effects balance out on average to confer no benefit relative to the unencoded case.

% Also shown, for comparison, are the input-state-independent unencoded remote CNOT gate results and a naive, first-order approximation to $F_{\mathrm{out}}$ in which only the bit flip error is considered. As three Werner states are used and there are no local errors, for this naive approximation, we approximate the probability of no error occurring as the probability that no entanglement error has occurred on any ebit, $F_{\mathrm{w}}^3$.

The natural next question to ask is for which input states is $F_{\mathrm{out}}$ better and worse than the unencoded case but there does not seem to be a discernible pattern to this behaviour and so we can only conclude that the results vary greatly with the input state.

\FloatBarrier

\bibliography{references}

\end{document}